\documentclass{article}
\usepackage{graphicx}  % standard LaTeX graphics tool;
\usepackage{amsmath}   % For getting proper math eqns
\usepackage{amssymb}   % Used for getting various symbols eg. gtrsim, lesssim etc.
\usepackage{bm} % For bold math (esp of lower greek letters)
\usepackage{dcolumn}% Align table columns on decimal point
\usepackage{color}
\usepackage{mathrsfs}
\usepackage{amsfonts}
\usepackage{varioref}
\usepackage[margin=1.6in]{geometry}
\usepackage{slashed}
\RequirePackage[colorlinks,citecolor=blue,urlcolor=magenta,linkcolor=blue]{hyperref}
\def\p{\partial}
\def\e{\epsilon}

\def\be{\begin{equation}}
\def\ee{\end{equation}}

\labelformat{section}{Section #1} 
\labelformat{subsection}{Section #1} 
\labelformat{subsubsection}{Section #1}
\labelformat{subsubsubsection}{Section #1}
\labelformat{equation}{Eq.~(#1)} 
\labelformat{figure}{Fig.~#1} 
\labelformat{subfigure}{Fig.~\thefigure#1} 
\labelformat{table}{Tab.~#1} 
\labelformat{appendix}{Appendix #1}
\title{\bf Non-perturbative $\langle \phi \rangle$, $\langle \phi^2 \rangle$ and the dynamically generated scalar mass with Yukawa interaction in the inflationary de Sitter spacetime }
\author{$^{1}$Sourav Bhattacharya\footnote{sbhatta.physics@jadavpuruniversity.in}\,\, and $^2$Moutushi Dutta Choudhury\footnote{mdchoudhury@kol.amity.edu}\\
\small{$^1$Relativity and Cosmology Research Centre, Department of Physics, Jadavpur University, Kolkata 700 032, India}\\
\small{$^2$Department of Physics, Amity Institute of Applied Sciences, Amity University\,-\,Kolkata,  AA2, Newtown, Kolkata 700 135, India}\\}
\begin{document}
\maketitle
%%%%%%%%%%%%%%%%%%%%%%%%%%%%%%%%%%%%%%%%%%%%%%%%%%%%%%%%%%%%%%%
\begin{abstract}
\noindent
We consider a massless minimally coupled self interacting quantum scalar field coupled to  fermion via the Yukawa interaction, in the inflationary de Sitter background. The fermion is also taken to be massless and the scalar potential is taken to be a hybrid, $V(\phi)= \lambda \phi^4/4!+ \beta \phi^3/3!$ ($\lambda >0$).    The  chief physical motivation behind this choice  of $V(\phi)$ corresponds to, apart from its boundedness from below property, the fact that shape wise $V(\phi)$ has qualitative similarity with standard inflationary classical slow roll  potentials. Also, its vacuum expectation value can be negative, suggesting some screening of the inflationary cosmological constant. We choose that $\langle \phi \rangle\sim 0$ at early times with respect to the Bunch-Davies vacuum, so that perturbation theory is valid initially.  We consider the equations satisfied by $\langle \phi (t) \rangle$   and  $\langle \phi^2(t) \rangle$, constructed from the coarse grained equation of motion for the slowly rolling $\phi$. We then compute the  vacuum diagrammes of various relevant operators using the in-in  formalism up to three loop, in terms of the leading powers of the secular logarithms. For a closed fermion loop, we have restricted ourselves here to only the local contribution. These large temporal logarithms are then resummed by constructing suitable non-perturbative equations to compute  $\langle \phi \rangle$   and  $\langle \phi^2 \rangle$.  $\langle \phi \rangle$ turns out to be at least approximately an order of magnitude less compared  to the minimum of the classical potential, $-3\beta/\lambda$, owing to the strong quantum fluctuations. For $\langle \phi^2 \rangle$, we have  computed the dynamically generated scalar mass at late times, by taking the appropriate purely local contributions. Variations of these quantities with respect to different couplings have also been presented.  
\end{abstract}
\vskip .5cm

\newpage

\tableofcontents

%%%%%%%%%%%%%%%%%%%%%%%%
\section{Introduction}\label{S1}
%%%%%%%%%%%%%%%%%%%%%%%

It is well accepted by now that the so called hot big bang cosmological model has been amply successful in explaining  the redshifts of galaxies, the origin of the cosmic microwave background radiation, abundance of light elements and the formation of  large scale cosmic structures of our universe. However,  a few puzzling issues such as the spatial flatness of our universe, the horizon problem and the hitherto unobserved relics such as the magnetic monopoles in terms of their scarcity,  cannot be addressed by this model~\cite{Rindler, Dicke, Wein, Mukhanov:2005sc}. The paradigm of the primordial cosmic inflation is a conjectured phase of very rapid, near exponential accelerated expansion of our very early universe,  introduced to provide possible answers  to these aforementioned  problems. In other words, the primordial inflation can be thought of as an initial condition to our universe. Indeed, inflation does not only solve these problems, but also provides an elegant mechanism  to generate primordial quantum cosmological density perturbations, as a seed to the large scale cosmic structures we observe today in the sky, see~\cite{Wein, Mukhanov:2005sc} and references therein for various theoretical and observational aspects of cosmic inflation. 

In order to drive an accelerated expansion of the universe, usually  some exotic matter with positive energy density but negative isotropic pressure is needed, known as the dark energy, the simplest form of which is regarded as the positive cosmological constant, $\Lambda$. For the latter in particular, the corresponding spacetime in the absence of any other backreaction is known as the de Sitter spacetime, which has an exponential scale factor and a maximal isometry group. Many computations can be done exactly in this background, owing to its maximal symmetry.   The early inflationary phase of our universe was expected to be dominated by dark energy/positive cosmological constant, whose density must had been much larger compared to that of we observe today.  It is an important task to understand how the inflationary $\Lambda$ got diminished to reach its current observed tiny value. This question seems to become more intriguing once we recall that only about $10\%$ deviation to the current $\Lambda$-value would have made the universe very different from what we observe it today, e.g.~\cite{Tsamis}. Since we are essentially talking about a time dependent, expanding spacetime background here, we must not ignore the  quantum effects which one may reasonably expect to be large in this context. At least a part of the problem then essentially boils down to estimating the backreaction of quantum fields,  to see to what extent these backreactions can screen the inflationary $\Lambda$ or break the de Sitter invariance, at the time of the graceful exit, see~\cite{Tsamis, Ringeval, Miao:2021gic} and references therein. We also refer our reader to  e.g.~\cite{Inagaki:2005qp, Dadhich, Padmanabhan, Alberte, Appleby, Khan:2022bxs, Evnin:2018zeo} for alternative proposals to the solution of the aforementioned cosmic coincidence as well as the cosmological constant problem. In any case, analysing matter field's backreaction in the inflationary background seems to be an important and challenging task.

Massless yet conformally non-invariant matter fields such as a massless minimally coupled  scalar or gravitons break the de Sitter invariance and can generate non-perturbative effects monotonically growing with time, popularly regarded as the {\it secular effect}, at late cosmological times~\cite{Floratos}.  These effects correspond to the late time, long wavelength infrared modes. There can be  instances in which such  infrared contributions can be resummed, to produce finite and physically acceptable  answers. The problem of resumming the secular effect for infrared gravitons however, remains as an open challenge, see~\cite{Miao:2021gic} and references therein. The secular effect for a self interacting massless and minimally coupled scalar field coupled to fermions in the de Sitter spacetime will be  the chief concern of this paper. 

Quantisation of the   massless or massive or conformally invariant scalar fields and their comoving Unruh-DeWitt detector responses in the de Sitter spacetime can be seen in~\cite{Chernikov:1968zm, Bunch:1978yq, Linde:1982uu, Starobinsky:1982ee, Allen:1985ux, Allen, Karakaya:2017evp, Ali:2020gij}. For a massless and  minimally coupled scalar, there exists no de Sitter invariant vacuum state, as was pointed out long ago  via the computation of the Wightman functions in~\cite{Allen:1985ux, Allen}. The corresponding de Sitter breaking term is given via the logarithm of the scale factor.  Consequently, while doing perturbation theory using these scalar two point functions, each internal line in a given Feynman diagram may yield one such logarithm, growing monotonically with time. Such temporal growth would eventually make the  amplitude  non-perturbatively large,  indicating a possible breakdown of the  perturbation theory after sufficient number of $e$-foldings. Such secular effects at one and two loops for self interacting scalar field theories, scalar quantum electrodynamics and even in the context of perturbative quantum gravity were computed in~\cite{Onemli:2002hr, Brunier:2004sb, Kahya:2009sz, Boyanovsky:2012qs, Onemli:2015pma, Prokopec:2007ak,  Liao:2018sci,  Miao:2020zeh, Glavan:2019uni, Karakaya:2019vwg, Cabrer:2007xm,  Boran:2017fsx, Akhmedov1, Akhmedov2} and references therein.  Attempts  to resum such non-perturbative infrared effects in order to extract finite quantities such as the dynamically generated mass  at late times can be seen in e.g.,~\cite{ Miao:2021gic, Moreau:2018ena, Moreau:2018lmz, Gautier:2015pca, Serreau:2013eoa, Serreau:2013koa, Serreau:2013psa, Ferreira:2017ogo, Burgess:2009bs, Burgess:2015ajz, Youssef:2013by, Baumgart:2019clc, Kitamoto:2018dek, Kamenshchik:2020yyn, Kamenshchik:2021tjh, Bhattacharya:2022aqi, Bhattacharya:2022wjl, Prokopec:2003tm} (also references therein). Note that such resummation should not be regarded as physically similar to that of the standard renormalisation group, as the secular effect is essentially a long wavelength infrared phenomenon and the renormalisation counterterms play no role while summing them, e.g.~\cite{Miao:2021gic}. 
\begin{figure}[h!]
\begin{center}
  \includegraphics[width=8.0cm]{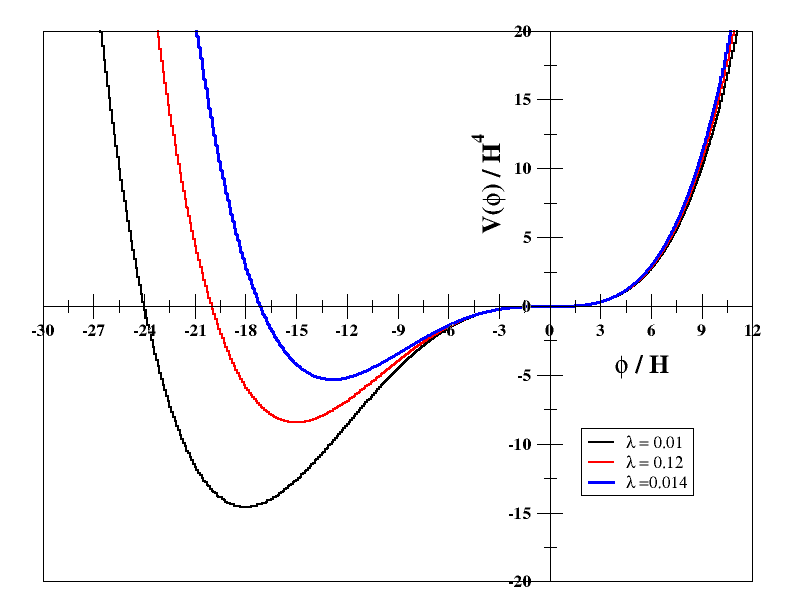}
 \caption{ \it \small The feature of the self interaction potential for the scalar field, \ref{y0-}, with $\bar{\beta}=\beta/H\sim 0.06$. All relevant quantities are made dimensionless with respect to the de Sitter Hubble constant, $H$. For negative values of $\bar{\beta}$, the minima just gets shifted to the other side of the vertical axis. Although the minima gets deepened with increasing $\bar{\beta}/\lambda$, the potential always remains bounded from below. We also note the shape-wise similarity of this hybrid potential to the standard slow roll (e.g.~\cite{Mukhanov:2005sc}) ones.   }
  \label{f0}
\end{center}
\end{figure}
A very efficient and infrared effective method to treat  a self interacting scalar field theory  non-perturbatively in the inflationary background   is the stochastic formalism, proposed in the pioneering works~\cite{Starobinsky:1986fx, Starobinsky:1994bd}.  We further refer our reader to~\cite{Cho:2015pwa, Prokopec:2015owa, Garbrecht:2013coa, Vennin:2015hra, Cruces:2022imf, Finelli:2008zg, Markkanen:2019kpv, Markkanen:2020bfc, Tsamis:2005hd} and references therein for recent developments. This  method is very efficient for computing expectation values associated with the super-Hubble or infrared part of the scalar field with a potential bounded from below, with respect to some late time equilibrium state. These deep infrared modes receive  quantum-kicks  from the stochastic forces rendered by the sub-Hubble modes as well as a drag due to the gradient of the potential, similar to that of the Brownian motion. The stochastic formalism basically maps a problem of quantum field theory into a classical statistical one. Putting things together now, all these non-perturbative results strongly suggest that it is the perturbation theory that remains invalid at late times, and proper non-perturbative treatment shows consistency with the de Sitter symmetry, owing to the dynamical mass generation of the scalar. A massive scalar, no matter with how much tiny mass, does not break de Sitter invariance. Computing this dynamically generated  mass and the late time back reaction of the scalar energy-momentum tensor {\it non-perturbatively} thus are important tasks for any given model, in order to understand how much screening of the inflationary $\Lambda$ is possible.\\

\noindent
In this paper we shall investigate the non-perturbative secular effect of a scalar field interacting with fermions via the Yukawa interaction, $g\bar{\psi}\psi \phi$, via quantum field theory.  In addition, the scalar is assumed   to be endowed with an asymmetric self interaction~\cite{Bhattacharya:2022aqi, Bhattacharya:2022wjl},  
\be
V(\phi)= \frac{\lambda \phi^4}{4!} +\frac{\beta \phi^3}{3!} \,\,\,\,\,\,(\lambda >0),
\label{y0-}
\ee
depicted in \ref{f0}.
While a massless minimal scalar with a quartic  self interaction in de Sitter spacetime is much well studied, to the best of our knowledge the above hybrid model of potential has not yet been investigated in detail. The physical motivation behind our choice of \ref{y0-} is as follows. First, we note the shape wise qualitative similarity of $V(\phi)$ with that of the standard slow roll ones. Also, $V(\phi)$ is renormalisable in four spacetime dimensions, making it apt to quantum field theory calculations.  Moreover, since $V(\phi)$ is bounded from below we expect the theory to be free from any runaway disaster. We shall assume that the system is located on the flat plateau ($\phi \sim 0$) of the potential {\it initially}, in the Bunch-Davies vacuum state. This also means $\phi$ has vanishing vacuum expectation value at the beginning. As time goes on, the system will roll down towards the minimum of the potential. However, due to strong quantum effects the potential will also receive non-trivial radiative corrections. Indeed, it was shown in~\cite{Bhattacharya:2022aqi} that the non-perturbative expectation value of $\phi$ at late times with respect to the initial Bunch-Davies vacuum is approximately one order of magnitude less compared to the position of the classical minima, $\phi=-3\beta/\lambda$, of $V(\phi)$. 
Also note  from \ref{y0-} that the system might generate negative vacuum expectation values at late times, for both $\phi$ and $V(\phi)$, effectively leading to some screening of the inflationary $\Lambda$~\cite{Bhattacharya:2022aqi, Bhattacharya:2022wjl}.  The present paper is an extension of these two earlier works, in the presence of fermions and Yukawa coupling.

Some of the earlier works  on the Yukawa theory in the de Sitter background can be seen 
in, e.g~\cite{Prokopec:2003qd, Duffy:2005ue, Miao:2006pn, Garbrecht, Boyanovski, Bhattacharya:2023twz}. These works chiefly involve computation of scalar and fermion self energies, renormalisation and some aspects of quantum entanglement.  In particular 
in~\cite{Miao:2006pn}, one loop effective action for the scalar was constructed by integrating out the fermions. However, any self interaction at tree level for the scalar was excluded there.  We further refer our reader to \cite{Toms:2018wpy, Toms:2018oal, Toms:2019erd, Inagaki:1993ya, Inagaki:1997kz, Inagaki:2015eza, Elizalde:1995bm} for discussion on effective action with Yukawa and four fermion interactions in  curved spacetimes. 
 
 The rest of the paper is organised is as follows. In the next section and in the two appendices, we sketch the basic technical framework we shall be working in. In \ref{s3}, we consider the late time, coarse grained equation of motion for the scalar and compute the late time expectation value of $\phi$ with respect to the initial Bunch-Davies vacuum up to three loop order.  In \ref{NP1}, we do resummation to find out a non-perturbative result. In \ref{s4}, we perform analogous analysis for $\phi^2$ and the dynamical generated late time mass is computed in \ref{loc-comp}. Finally we conclude in \ref{s5}. We shall work with the mostly positive signature of the metric in $d = 4 - \epsilon$ ($\epsilon = 0^{+}$) dimensions and will set $c = 1 = \hbar$ throughout. The vacuum expectation value of any operator $O$ will be denoted as $\langle O \rangle$. Also for the sake of brevity and to save space, we shall denote for powers of propagators and logarithms respectively as, $(i\Delta)^n\equiv i\Delta^n$ and also $(\ln a )^n \equiv \ln^n a$.

%%%%%%%%%%%%%
\section{The basic set up}\label{loop1}
%%%%%%%%%%%%%
The metric for the inflationary de Sitter spacetime reads respectively in the cosmological and conformal temporal coordinates
\begin{eqnarray}
ds^2 = -dt^2 + a^2(t) d{\vec x}^2= a^2(\eta) \left(-d\eta^2 +  d{\vec x}^2\right)
\label{y0}
\end{eqnarray}
where $a(t)=e^{Ht}$ or $a(\eta)=  -1/H\eta$ is the de Sitter scale factor and $H=\sqrt{\Lambda/3}$ is the de Sitter Hubble rate.  We have the range $0 \leq t < \infty$, so that $-H^{-1}\leq \eta <0^-$. Note that the temporal level of the initial hypersurface can easily 
be achieved as we wish, by exploiting the time translation symmetry of the de Sitter spacetime.

The bare Lagrangian density corresponding for the matter sector reads,
\begin{eqnarray}
\sqrt{-g}{\cal L}= \sqrt{-g} \left[ -\frac12 g^{\mu\nu} (\nabla_{\mu} \psi')(\nabla_{\nu} \psi') -\frac12 m_0^2 \psi'^2 - \frac{\lambda_0}{4!} \psi'^4 - \frac{\beta_0}{3!} \psi'^3 - \tau_0 \psi' + i\overline{\Psi}\slashed{\nabla} \Psi - m_{0 f} \overline{\Psi} \Psi -g_0 \overline{\Psi} \Psi \psi   \right]
\label{y1}
\end{eqnarray}
where $\slashed{\nabla}= \gamma^{\mu}\nabla_{\mu}$, with   $\nabla_{\mu}$ being the spin covariant derivative. Also, since we are working with the mostly positive signature of the metric, we shall take the anti-commutation relation for the $\gamma$-matrices as
\be
[\gamma_{\mu},\gamma_{\nu}]_+= -2 g_{\mu\nu} \, {\bf I}_{d\times d}
\label{Ac}
\ee

Defining the field strength renormalisation, $\phi=\psi'/\sqrt{Z_b}$ and $\psi=\Psi/\sqrt{Z_f}$, we have  in $d$-dimensions
\begin{eqnarray}
\sqrt{-g}{\cal L}=  -\frac{Z_b}{2} \eta^{\mu\nu} a^{d-2} (\p_{\mu} \phi)(\p_{\nu} \phi) -\frac{1}{2} Z_b m_0^2 \phi^2 a^d - \frac{Z_b^2 \lambda_0}{4!} \phi^4 a^d - \frac{\beta_0 Z_b^{3/2}}{3!} \phi^3 a^d - \tau_0 \sqrt{Z_b} \phi \,a^d \nonumber\\- i Z_f\overline{\psi}\slashed{\nabla} \psi \,a^d - m_{0 f} Z_f  \overline{\psi} \psi \,a^d -g_0 Z_f \sqrt{Z_b}\, \overline{\psi} \psi \phi\, a^d 
\label{y2}
\end{eqnarray}
We take both scalar and fermion rest masses to be vanishing. We write
\begin{eqnarray}
&&Z_b= 1+\delta Z \qquad Z m_0^2 = 0+\delta m^2 \qquad Z^2 \lambda_0 = \lambda +\delta \lambda \qquad \beta_0 Z^{3/2}= \beta+\delta \beta   \nonumber\\ && \tau_0 \sqrt{Z} =\alpha \qquad 
Z_f =1+\delta Z_f \qquad m_{0f} Z_f= 0+\delta m_f \qquad g_0 Z_{f}\sqrt{Z}= g+\delta g  
\label{y3}
\end{eqnarray}

The above decomposition splits \ref{y2} as
\begin{eqnarray}
&&\sqrt{-g}{\cal L}=  -\frac{1}{2} \eta^{\mu\nu} a^{d-2} (\p_{\mu} \phi)(\p_{\nu} \phi)  - \frac{\lambda}{4!} \phi^4 a^d - \frac{\beta}{3!} \phi^3 a^d - \alpha \phi a^d + i \overline{\psi}\slashed{\nabla} \psi \,a^d  -g\, \overline{\psi} \psi \phi\, a^d \nonumber\\&&-\frac{\delta Z}{2} \eta^{\mu\nu} a^{d-2} (\p_{\mu} \phi)(\p_{\nu} \phi)  - \frac12 \delta m^2 \phi^2-\frac{\delta \lambda}{4!} \phi^4 a^d - \frac{\delta\beta}{3!} \phi^3 a^d  + i Z_f \overline{\psi}\slashed{\nabla} \psi \,a^d  -\delta m_f \overline{\Psi}\Psi a^d - \delta g\, \overline{\psi} \psi \phi\, a^d 
\label{y4}
\end{eqnarray}

The propagator for a massless and minimally coupled scalar field in the de Sitter background reads~\cite{Onemli:2002hr},
\be
i\Delta(x,x')= A(x,x')+ B(x,x')+ C(x,x')
\label{props1}
\ee
where 
\begin{eqnarray}
&&A(x,x') = \frac{H^{2-\e} \Gamma(1-\e/2)}{4\pi^{2-\e/2}}\frac{1}{y^{1-\e/2}} \nonumber\\
&&B(x,x') =  \frac{H^{2-\e} }{(4\pi)^{2-\e/2}}\left[-\frac{2\Gamma(3-\e)}{\e}\left(\frac{y}{4} \right)^{\e/2}+ \frac{2\Gamma(3-\e)}{\e\,\Gamma(2-\e/2)} + \frac{2\Gamma(3-\e)}{\Gamma(2-\e/2)}\ln (aa')\right]\nonumber\\
&&C(x,x')= \frac{H^{2-\e} }{(4\pi)^{2-\e/2}} \sum_{n=1}^{\infty} \left[\frac{\Gamma(3-\e+n)}{n\Gamma(2-\e/2+n)}\left(\frac{y}{4} \right)^n- \frac{\Gamma(3-\e/2+n)}{(n+\e/2)\Gamma(2+n)}\left(\frac{y}{4} \right)^{n+\e/2}  \right]
\label{props2}
\end{eqnarray}
where the de Sitter invariant biscalar interval reads
\be
y(x,x')= aa'H^2 \Delta x^2 = aa'H^2 \left[ |\vec{x}-\vec{x'}|^2- (\eta-\eta')^2\right]
\label{props3}
\ee
where we have abbreviated $a\equiv a(\eta)$ and $a'\equiv a(\eta')$. There are four propagators pertaining to the in-in or the Schwinger-Keldysh formalism we shall be using, briefly outlined in \ref{A}, characterised by suitable four complexified  distance functions, $\Delta x^2$,
\begin{eqnarray}
&&\Delta x^2_{++} =\left[ |\vec{x}-\vec{x'}|^2- (|\eta-\eta'|-i\e)^2\right]= (\Delta x^2_{--})^{*}\nonumber\\
&&\Delta x^2_{+-} =\left[ |\vec{x}-\vec{x'}|^2- ((\eta-\eta')+i\e)^2\right]= (\Delta x^2_{-+})^{*} \qquad (\e=0^+)
\label{props3}
\end{eqnarray}
The first two correspond respectively to the Feynman and anti-Feynman propagators, whereas the last two correspond to the two Wightman functions.

From \ref{props2}, we have in the coincidence limit for all the four propagators
\be 
i\Delta(x,x) = \frac{H^{2-\e} \Gamma(2-\e)}{2^{2-\e} \pi^{2-\e/2} \Gamma(1-\e/2)}\left(\frac{1}{\e}+\ln a  \right)
\label{y6}
\ee

Using the above expression, we may compute the one loop tadpole (corresponding to the cubic self interaction) and the self energy bubble (corresponding to the quartic self interaction) diagrams.   The corresponding  one loop  renormalisation counterterms read
\be 
\alpha= -\frac{\beta H^{2-\e} \Gamma(2-\e)}{2^{3-\e}\pi^{2-\e/2}\Gamma(1-\e/2)\e} \qquad \delta m_{\lambda}^2 = - \frac{\lambda H^{2-\e} \Gamma(2-\e)}{2^{3-\e} \pi^{2-\e/2} \Gamma(1-\e/2)\e}
\label{y7}
\ee

The square of the scalar propagator reads~\cite{Brunier:2004sb}
\begin{eqnarray}
i\Delta_{++}^2(x,x')= -\frac{i\mu^{-\e} a'^{-4+2\e}\Gamma(1-\e/2)}{2^3\pi^{2-\e/2}\e (1-\e)}\delta^d(x-x') \,+ \frac{H^4}{2^6 \pi^4} \ln^2 \frac{\sqrt{e}H^2 \Delta x^2_{++}}{4} - \frac{H^2 (aa')^{-1}}{2^4\pi^4} \frac{\ln \frac{\sqrt{e}H^2 \Delta x^2_{++}}{4}}{\Delta x^2_{++}},
\label{e51'}
\end{eqnarray} 
where $\mu$ is the renormalisation scale.  This leads to the one loop mass renormalisation counterterm for the cubic sector
\begin{eqnarray}
\delta m_{\beta}^2= \frac{\beta^2 \mu^{-\e} \Gamma(1-\e/2)}{2^4 \pi^{2-\e/2} \e (1-\e)}
\label{e51'add}
\end{eqnarray} 

Since the de Sitter spacetime \ref{y0} is conformally flat, and a massless fermion is conformally invariant, the propagator for the same corresponds to just the Minkowski space propagator multiplied with appropriate powers of the scale factor, e.g.~\cite{Duffy:2005ue}
\be
iS(x,x') = \frac{\Gamma(d/2-1)}{4\pi^{d/2}(aa')^{(d-1)/2}} i\slashed{\p}\frac{1}{(\Delta x^2)^{d/2-1}}= -\frac{i(d-2)\Gamma(d/2-1)}{4\pi^{d/2} (aa')^{(d-1)/2}} \frac{\gamma_{\mu}\Delta x^{\mu}}{(\Delta x^2)^{d/2}}
\label{y10}
\ee
There are  four fermion propagators, $iS_{\pm\pm}(x,x')$, corresponding to the four distance functions defined in \ref{props3} in the in-in formalism. Taking now the trace of the square of the above propagator using \ref{Ac}, we have
\be
{\rm Tr}(iS^2(x,x')) = \frac{(d-2)^2\Gamma^2(d/2-1)}{4\pi^{d}(aa')^{d-1}}\frac{1}{\Delta x^{2(d-1)}} 
\label{y11}
\ee
Recall that in our notation,  $(iS^2(x,x')) \equiv (iS(x,x'))^2 $ (cf., the end of the preceding section).  Using 
$$\frac{1}{\Delta x^{2(d-1)}}= \frac{1}{2(d-2)^2} \p^2 \frac{1}{\Delta x^{2(d-2)}}= \frac{1}{4(d-2)^2(d-3)(d-4)} \p^4 \frac{1}{\Delta x^{2(d-3)}}$$
and substituting it into \ref{y11}, we have
\begin{eqnarray}
{\rm Tr}(iS_{++}^2(x,x')) =&& -\frac{i\mu^{-\e} \Gamma(1-\e/2) }{2^2\pi^{2-\e/2}\e(1-\e)(aa')^{3-\e/2}}\p^2 \delta^d(x-x')-  \frac{i \mu^{-\e}\ln(aa')}{2^3\pi^2(aa')^3}\p^2 \delta^4(x-x')- \frac{3i\e \mu^{-\e}\ln^2(aa')}{2^5\pi^2(aa')^3}\p^2 \delta^4(x-x')\nonumber\\&& - \frac{1}{2^8\pi^4 (aa')^3}\p^6\left( \ln^2 \Delta x_{++}^2-2\ln \Delta x^2_{++}\right)+{\cal O}(\e^2)
\label{y12}
\end{eqnarray}
${\rm Tr}(iS_{--}^2(x,x'))$ is given by the complex conjugation of the above result. ${\rm Tr}(iS_{+-}^2(x,x'))$ and ${\rm Tr}(iS_{-+}^2(x,x'))$ are complex conjugate to each other, with the local terms containing the $\delta$-function are absent. 

Let us now consider the one loop self energy for scalar due to the Yukawa interaction.  It reads 
\begin{eqnarray}
g^2 (aa')^d {\rm Tr} \left(iS(x',x) i S(x,x') \right) =  -g^2 (aa')^d {\rm Tr}\,  i S^2(x,x') 
\label{y12'}
\end{eqnarray}
The divergence in the self energy can be absorbed by scalar field strength renormalisation counterterm and a conformal counterterm, such that their combination introduces a term in the Lagrangian density~\cite{Duffy:2005ue}  
$$  (aa')^{d/2} \left( \frac{\delta Z}{aa'} \p^2 \delta^d (x-x')\right),  $$
whose contribution, when added to the self energy, gives us the scalar field strength renormalisation counterterm 
\be
\delta Z = -\frac{ g^2 \mu^{-\e} \Gamma(1-\e/2) }{2^2\pi^{2-\e/2}\e(1-\e)}
\label{y13}
\ee

The above expression for the scalar self energy due to the fermion loop will be essential to compute various diagrams for our purpose. We note  that the last term on the right hand side of \ref{y12} is non-local. However, since it contains six derivatives, and we are interested only in the leading late time secular logarithms, we may safely ignore it, and focus only on the terms containing the $\delta$-functions.  In other words, we shall only be interested in the local part of the contribution coming from a closed  fermion loop. This also means that we shall not require the fermion Wightman functions $iS_{+-}(x,x') $ or $iS_{-+}(x,x')$ for our present purpose.

%%%%%%%%%%%%%%
\section{Perturbative computation of $\langle \phi \rangle$ and its resummation}\label{s3}
%%%%%%%%

 We begin with the full equation of motion for the scalar field
\begin{eqnarray}
\Box \phi -\frac{\lambda \phi^3}{6} -\frac{\beta \phi^2}{2} - g \overline{\Psi}\Psi
=0,
\label{e1}
\end{eqnarray}
and take its expectation value with respect to the initial Bunch-Davies vacuum in the in-in formalism. By the symmetry of the background de Sitter space, we expect that $\langle \phi \rangle$ will be spatially homogeneous. Apart from this, under the slow roll condition {\it and} at sufficiently late times or super-Hubble limit, we expect that the second temporal derivative of $\langle \phi \rangle$ will be subleading compared to that of the first derivative. In this context we may also recall that the perturbative $\langle \phi \rangle$ at each order at sufficiently late times has a generic, leading secular behaviour$\sim {\cal N}^n$, where ${\cal N}=Ht$ is the number of de Sitter $e$-foldings and $n\geq 2$ is a positive integer~\cite{Bhattacharya:2022aqi}. Thus the second temporal derivative of $\langle \phi \rangle$ will indeed be subleading compared to that of the first derivative, for ${\cal N} \gg 1$ at each perturbative order. Putting these in together, we have the  equation of motion satisfied by the slowly rolling $\langle \phi \rangle$ in the super-Hubble limit,  
\begin{eqnarray}
\frac{d \langle {\phi}\rangle}{d t} +\frac{\lambda}{18H}\langle \phi^3\rangle +\frac{\beta}{6H} \langle\phi^2\rangle + \frac{g}{3H} \langle\overline{\Psi}\Psi \rangle
=0
\label{e2}
\end{eqnarray}
Note that similar equation  can also be found in the stochastic formalism, e.g.~\cite{Fujita:2014tja},  in which $\langle \phi \rangle$ basically corresponds to the spatially homogeneous  field, coarse grained over a volume larger than the Hubble radius. Also, $\langle \phi \rangle$ found by integrating  \ref{e2} matches exactly with the one found directly from the tadpoles,  as shown in \ref{C}  for the sake of consistency, for the scalar self interaction. Let us now compute  the un-differentiated  vacuum expectation values appearing in \ref{e2}.

%%%%%%%
\subsection{Computation of $\langle \phi^3 \rangle$}\label{SSphi3}
%%%%%%%%%
%
\begin{figure}[h!]
\begin{center}
  \includegraphics[width=12.0cm]{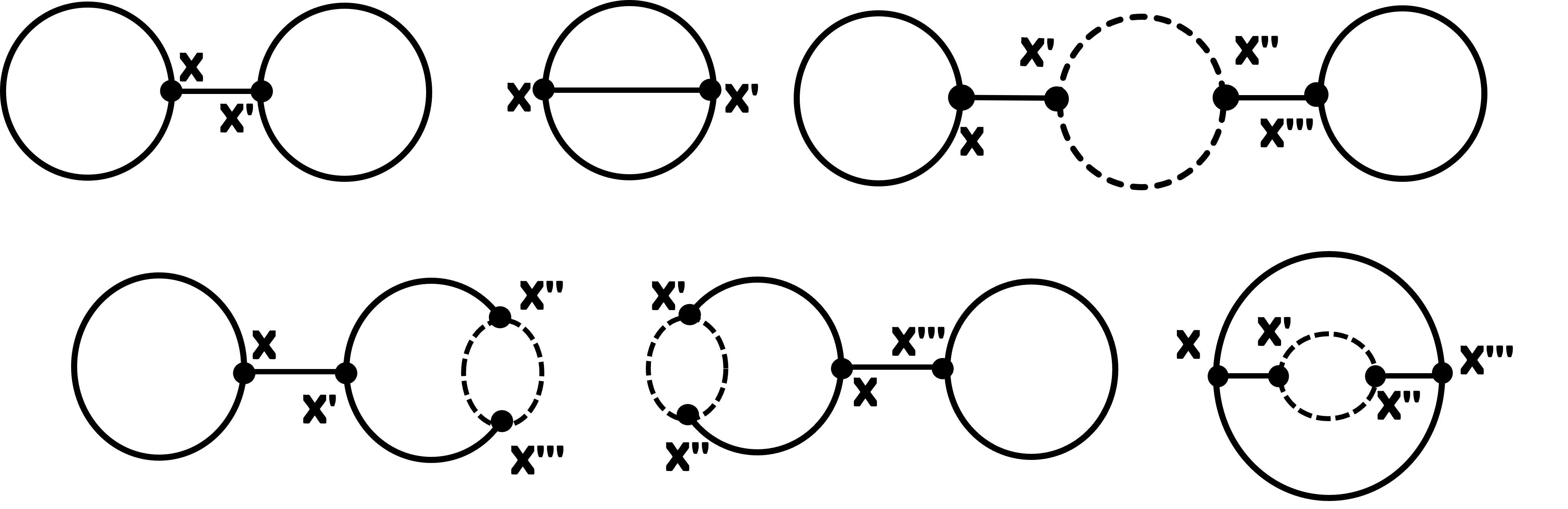}
 \caption{\small \it Vacuum diagrams for perturbative computation of $\langle \phi^3 (x)\rangle$ up to three loop. Solid and dashed lines represent respectively the scalar and the fermion.  All primed vertices  are integrated over. }
  \label{f2'}
\end{center}
\end{figure}

In order to compute  $\langle \phi^3\rangle$ perturbatively, we shall be dealing with   six diagrams at ${\cal O}(\beta)$ and ${\cal O}( \beta g^2 )$  as shown in \ref{f2'}. 
The first diagram in the in-in or Schwinger-Keldysh formalism outlined in \ref{A} reads 
\begin{eqnarray}
-\frac{3i\beta}{2} \int d^d x' a'^d i\Delta(x,x) i\Delta (x',x') \left(i\Delta_{++}(x,x')-i\Delta_{+-}(x,x') \right)
\label{e3}
\end{eqnarray}
Note that the tadpoles of this diagram give divergences from  \ref{y6}, which can be canceled  via the one loop tadpole  counterterm \ref{y7}.  However, 
we really do  not need to  bother ourselves with renormalisation for now, as we are essentially working in a late time long wavelength framework.  We shall discuss some renormalisation instead in \ref{loc-comp}, where we  need to compute some purely local contributions, in order to compute the dynamically generated mass. The leading secular contribution from any diagram is essentially a non-local one free from any ultraviolet divergences, giving one secular logarithm for {\it each} internal line. The local contributions contain  $\delta$-functions and consist of  divergent plus finite parts. The finite part may contain secular logarithms as well but their power are less than the total number of internal lines. 

We  recall now that  $\eta$ is the final time, \ref{B}. Thus   from  \ref{propagatoridentities}, we have for  \ref{e3} 
\begin{eqnarray}
&&\frac{3i\beta}{2} \int d^d x' a'^d i\Delta(x,x) i\Delta (x',x') \left(i\Delta_{+-}(x,x')-i\Delta_{-+}(x,x') \right) \nonumber\\&&
=\frac{3i\beta H^4}{2^5\pi^4}\ln a\times  \int d\eta' a'^4  \ln a' \left(i\Delta_{+-}(k=0,\eta,\eta')- i\Delta_{-+}(k=0,\eta, \eta')  \right)
= -\frac{\beta H^2}{2^6 \pi^4}\ln^3 a \,+\,{\cal O}(\ln^2 a)
\label{e4}
\end{eqnarray}
where we have used the finite part of \ref{y6} and the ingredients of \ref{B}, derived using the spatial momentum space.  We  note that the three internal lines of the diagram has yielded the cubic power of the late time secular logarithm at the leading order. \\

\noindent
The in-in amplitude for the the second diagram of \ref{f2'} reads
\begin{eqnarray}
-i\beta \int d^4 x' a'^4 \left(i\Delta^3_{++}(x,x')-i\Delta^3_{+-}(x,x') \right)
\label{e5}
\end{eqnarray}
Using the expressions given towards the end of  \ref{A}, and the spatial momentum space as discussed in \ref{B}, we have after some algebra the leading late time result
\begin{eqnarray}
i\beta \int d^4 x' a'^4 \left(i\Delta^3_{+-}(x,x')-i\Delta^3_{-+}(x,x') \right)= -\frac{\beta H^2}{2^4 \times 3 \pi^4}\ln^3 a
\label{e6}
\end{eqnarray}

\noindent
The third of \ref{f2'} contains the Yukawa interaction. Recall that we shall work only with the local part of the fermion loop here (cf., the discussion towards the end of \ref{loop1}). This means we may ignore the self-energies due to the fermion Wightman functions, $iS_{\pm\mp}(x,x')$.  Thus the relevant contribution reads 
\begin{eqnarray}
&&3i\beta g^2 i\Delta(x,x) {\rm Tr} \int d^4 x' d^4 x'' d^4 x''' (a' a'' a''')^4 i\Delta(x''',x''') \left[i\Delta_{-+}(x,x') i\Delta_{++}(x'',x''') iS^2_{++}(x',x'') \right. \nonumber\\&& \left.+ i\Delta_{+-}(x,x') i\Delta_{-+}(x'',x''') iS^2_{--}(x',x'')- i\Delta_{-+}(x,x') i\Delta_{+-}(x'',x''') iS^2_{++}(x',x'')-i\Delta_{+-}(x,x') i\Delta_{--}(x'',x''') iS^2_{--}(x',x'') \right] \nonumber\\
\label{e7}
\end{eqnarray}
It is easy to see that the above integral is non-vanishing only for the temporal hierarchy $\eta', \eta'' \gtrsim \eta'''$. Using this and the renormalised expression for the scalar self energy due to fermion loop derived towards the end of \ref{loop1}, the above integral becomes  
\begin{eqnarray}
&&\frac{3\beta g^2 H^4}{2^7\pi^6} \ln a  \int d^4 x' d^4 x'' d^4 x''' (a' a'') a'''^4 \ln a'''  \ln (a' a'') (\p_{x''}^2 \delta^4(x'-x'')) \nonumber\\
&&\times   \left(i\Delta_{+-}(x,x')-i\Delta_{-+}(x,x') \right) \left(i\Delta_{+-}(x'',x''')-i\Delta_{-+}(x'',x''') \right) 
\label{e8}
\end{eqnarray}
Integrating \ref{e8} now by parts utilising the $\delta$-function, we have 
\begin{eqnarray}
&&\frac{3\beta g^2 H^4}{2^6\pi^6} \ln a  \int d^4 x' d^4 x''' a'''^4 \ln a'''\,\p^2_{x'}\left[ a'^2 \ln a' \times   \left(i\Delta_{+-}(x,x')-i\Delta_{-+}(x,x') \right) \left(i\Delta_{+-}(x',x''')-i\Delta_{-+}(x',x''') \right) \right]\nonumber\\
\label{e8'}
\end{eqnarray}

Let us first evaluate the spatial derivative part, $\p^2_{\vec{x}'}$. Introducing the $3$-momentum space as in \ref{B}, the relevant part of the above integration reads
\begin{eqnarray}
&&\int d^3 \vec{x}' d^3 \vec{x}''' \frac{d^3 \vec{k}_1d^3 \vec{k}_2 }{(2\pi)^6} \p^2_{\vec{x}'} \left[e^{i \vec{k}_1\cdot (\vec{x}- \vec{x'})}e^{i \vec{k}_2\cdot (\vec{x'}- \vec{x'''})} \left( i\Delta_{+-}(k_1, \eta, \eta')- i\Delta_{-+}(k_1, \eta, \eta')\right)\left(i\Delta_{+-}(k_2, \eta', \eta''')- i\Delta_{-+}(k_2, \eta', \eta''') \right)\right]
\nonumber\\
&&=-\int d^3 \vec{x}'  \frac{d^3 \vec{k}_1}{(2\pi)^3} \vec{k}^2_{1} e^{i \vec{k}_1\cdot (\vec{x}- \vec{x'})} \left( i\Delta_{+-}(k_1, \eta, \eta')- i\Delta_{-+}(k_1, \eta, \eta')\right)\left(i\Delta_{+-}(0, \eta', \eta''')- i\Delta_{-+}(0, \eta', \eta''') \right)
\end{eqnarray}
We recall that $\left( i\Delta_{+-}(k, \eta, \eta')- i\Delta_{-+}(k, \eta, \eta')\right)=i/3Ha'^3$ at the leading order,  irrespective of the value of $k$, \ref{B}. 
Performing now the $\vec{x'}$ integration gives a $\delta^3(\vec{k}_1)$, forcing thus the integration to vanish, giving us the relevant part of \ref{e8'},
\begin{eqnarray}
&&-\frac{3\beta g^2 H^4}{2^6\pi^6} \ln a  \int d^4 x' d^4 x''' a'''^4 \ln a'''\,\p^2_{\eta'}\left[ a'^2 \ln a' \times   \left(i\Delta_{+-}(x,x')-i\Delta_{-+}(x,x') \right) \left(i\Delta_{+-}(x',x''')-i\Delta_{-+}(x',x''') \right) \right]\nonumber\\
\label{e8add3}
\end{eqnarray}
Employing the spatial momentum space again, the above integral becomes
\begin{eqnarray}
&&-\frac{3\beta g^2 H^4}{2^6\pi^6} \ln a  \int d \eta' d \eta''' a'''^4 \ln a'''\,\p^2_{\eta'}\left[ a'^2 \ln a' \times   \left(i\Delta_{+-}(0, \eta,\eta')-i\Delta_{-+}(0,\eta,\eta') \right) \left(i\Delta_{+-}(0,\eta',\eta''')-i\Delta_{-+}(0,\eta',\eta''') \right) \right]\nonumber\\&&
=\frac{\beta g^2 H^2}{2^6\times 3\pi^6}\ln a \int  d \eta' d \eta''' a''' \ln a'''\p^2_{\eta'} \left(\frac{\ln a'}{a'}\right)=- \frac{\beta g^2 H^2}{2^6\times 3\pi^6}\ln a  \int \frac{ d a'}{a'} \frac{d a'''} {a'''} \ln a'''=-\frac{\beta g^2 H^2}{2^7\times 9\pi^6}\ln^4 a
\label{e8add4}
\end{eqnarray}
where in the last line we have used  the temporal hierarchy, $\eta' \gtrsim \eta'''$.\\

\noindent
Likewise, the fourth diagram of \ref{f2'} equals
\begin{eqnarray}
&&3i\beta g^2 i\Delta(x,x) {\rm Tr} \int d^4 x' d^4 x'' d^4 x''' (a' a'' a''')^4  \left[i\Delta_{-+}(x,x') i\Delta_{++}(x',x'')i\Delta_{++}(x',x''') iS^2_{++}(x'',x''') \right. \nonumber\\&& \left. +i\Delta_{-+}(x,x') i\Delta_{+-}(x',x'')i\Delta_{+-}(x',x''') iS^2_{--}(x'',x''') -i\Delta_{+-}(x,x') i\Delta_{-+}(x',x'')i\Delta_{-+}(x',x''') iS^2_{++}(x'',x''')\right. \nonumber\\&& \left.   -i\Delta_{+-}(x,x') i\Delta_{--}(x',x'')i\Delta_{--}(x',x''') iS^2_{--}(x'',x''')    \right]
\label{e9}
\end{eqnarray}

The above integral is non-vanishing for $\eta' \gtrsim \eta'', \eta'''$, hence can be written as 
\begin{eqnarray}
&& \frac{3\beta g^2 H^2}{2^5 \pi^4}\ln a \int d^4 x' d^4 x'' d^4 x''' a'^4(a''a''') \ln (a'' a''') (\p^2_{x'''} \delta^4 (x''-x''') )\nonumber\\
&&\times \left(i\Delta_{+-}(x,x')-i\Delta_{-+}(x,x') \right) \left(i\Delta_{+-}(x',x'')i\Delta_{+-}(x',x''')-i\Delta_{-+}(x',x'')i\Delta_{-+}(x',x''') \right) \nonumber\\
&&= -\frac{3\beta g^2 H^2}{2^4 \pi^4}\ln a \int d^4 x' d^4 x''  a'^4 \p^2_{x''}\left[  a''^2 \ln a''   \left(i\Delta_{+-}(x,x')-i\Delta_{-+}(x,x') \right) \left(i\Delta^2_{+-}(x',x'')-i\Delta^2_{-+}(x',x'') \right)\right]
\label{e10}
\end{eqnarray}

We evaluate the above as earlier. The spatial derivative part of the integral, $\p^2_{\vec{x''}}$ vanishes and we obtain by employing the spatial momentum space,
\be
-\frac{\beta g^2 H^2}{2^5 \times 9\pi^6} \ln^4 a
\label{e10add}
\ee

\noindent
The fifth diagram of \ref{f2'} equals,
\begin{eqnarray}
&&3i\beta g^2{\rm Tr} \int d^4 x' d^4 x'' d^4 x''' (a' a'' a''')^4 i\Delta (x',x') \left[ i\Delta_{-+}(x,x')i\Delta_{-+}(x,x'')i\Delta_{-+}(x,x''') iS^2_{++}(x'',x''') \right. \nonumber\\&&\left. +  i\Delta_{-+}(x,x')i\Delta_{+-}(x,x'')i\Delta_{+-}(x,x''') iS^2_{--}(x'',x''')- i\Delta_{+-}(x,x')i\Delta_{-+}(x,x'')i\Delta_{-+}(x,x''') iS^2_{++}(x'',x''') \right. \nonumber\\&&\left.- i\Delta_{+-}(x,x')i\Delta_{+-}(x,x'')i\Delta_{+-}(x,x''') iS^2_{--}(x'',x''')\right]
\nonumber\\
&&= \frac{3\beta g^2 H^2}{2^5 \pi^4}\int d^4 x' d^4 x'' d^4 x''' a'^4(a''a''')\ln a' \ln (a'' a''') (\p^2_{x'''} \delta^4 (x''-x''')) \nonumber\\
&&\times \left(i\Delta_{+-}(x,x')-i\Delta_{-+}(x,x') \right) \left(i\Delta_{+-}(x,x'')i\Delta_{+-}(x,x''')-i\Delta_{-+}(x,x'')i\Delta_{-+}(x,x''') \right) 
\label{e10add2}
\end{eqnarray}

The spatial derivative part vanishes as earlier to yield,
\begin{eqnarray}
-\frac{\beta g^2 H^2}{2^6 \times 3\pi^6} \ln^4 a
\label{e10add3}
\end{eqnarray}

\noindent
The sixth diagram of \ref{f2'} equals 
\begin{eqnarray}
&& 2i\beta g^2{\rm Tr}\int d^4 x' d^4 x'' d^4 x''' (a'a''a''')^4  \left[i\Delta^2_{-+}(x,x''')i\Delta_{-+}(x,x')i\Delta_{++}(x'',x''') iS^2_{++}(x',x'') \right. \nonumber\\&&\left.
+ i\Delta^2_{-+}(x,x''')i\Delta_{+-}(x,x')i\Delta_{-+}(x'',x''') iS^2_{--}(x',x'')  -i\Delta^2_{+-}(x,x''')i\Delta_{-+}(x,x')i\Delta_{+-}(x'',x''') iS^2_{++}(x',x'') \right. \nonumber\\&&\left.
 -i\Delta^2_{+-}(x,x''')i\Delta_{+-}(x,x')i\Delta_{--}(x'',x''') iS^2_{--}(x',x'') \right]
\label{e11}
\end{eqnarray}
For $\eta'' \gtrsim \eta'''$, this gives,
\begin{eqnarray}
&& \frac{\beta g^2}{2^2\pi^2}  \int d^4 x' d^4 x'' d^4 x''' (a'a'')a'''^4 \ln (a'a'') \p^2_{x''}\delta^4(x'-x'')\nonumber\\&&\times  \left(i\Delta_{+-}(x,x') -i\Delta_{-+}(x,x')\right) \left(i\Delta^2_{+-}(x,x''')i\Delta_{+-}(x'',x''') - i\Delta^2_{-+}(x,x''')i\Delta_{-+}(x'',x''')\right)
= -\frac{\beta g^2 H^2}{2^7 \times 9\pi^6} \ln^4a\nonumber\\
\label{e12}
\end{eqnarray}

For $\eta''' \gtrsim \eta''$, \ref{e11} becomes,
\begin{eqnarray}
&& \frac{\beta g^2}{2^2\pi^2}  \int d^4 x' d^4 x'' d^4 x''' (a'a'')a'''^4 \ln (a'a'') \p^2_{x''}\delta^4(x'-x'') \nonumber\\&&
    \times \left(i\Delta^2_{+-}(x,x''')   -i\Delta^2_{-+}(x,x''') \right) \left(i\Delta_{+-}(x,x') i\Delta_{+-}(x''',x'')-i\Delta_{-+}(x,x') i\Delta_{-+}(x''',x'') \right) = -\frac{\beta g^2 H^2}{2^6 \times 9\pi^6} \ln^4 a\nonumber\\
\label{e13}
\end{eqnarray}
where we have also used above $i\Delta_{-+}(x'',x''')=i\Delta_{+-}(x''',x'') $.\\

\noindent
Adding the above with \ref{e12}, the final result for \ref{e11} follows. Also, combining this with \ref{e4}, \ref{e6}, \ref{e8add4}, \ref{e10add} and \ref{e10add3}, we obtain the perturbative result for $\langle \phi^3\rangle$ in terms of the leading powers of the secular logarithms
\begin{eqnarray}
\langle \phi^3\rangle =  -\frac{\beta H^2}{2^6 \times 3\pi^4} 7\ln^3 a - \frac{\beta g^2 H^2}{2^6 \times 9\pi^6} 7\ln^4 a +\,{\cal O}(\lambda \beta g^2) 
\label{e13add}
\end{eqnarray}
%

%%%%%%%
\subsection{Computation of $\langle \phi^2 \rangle$}
%%%%%%%%
%
\begin{figure}[h!]
\begin{center}
  \includegraphics[width=9.0cm]{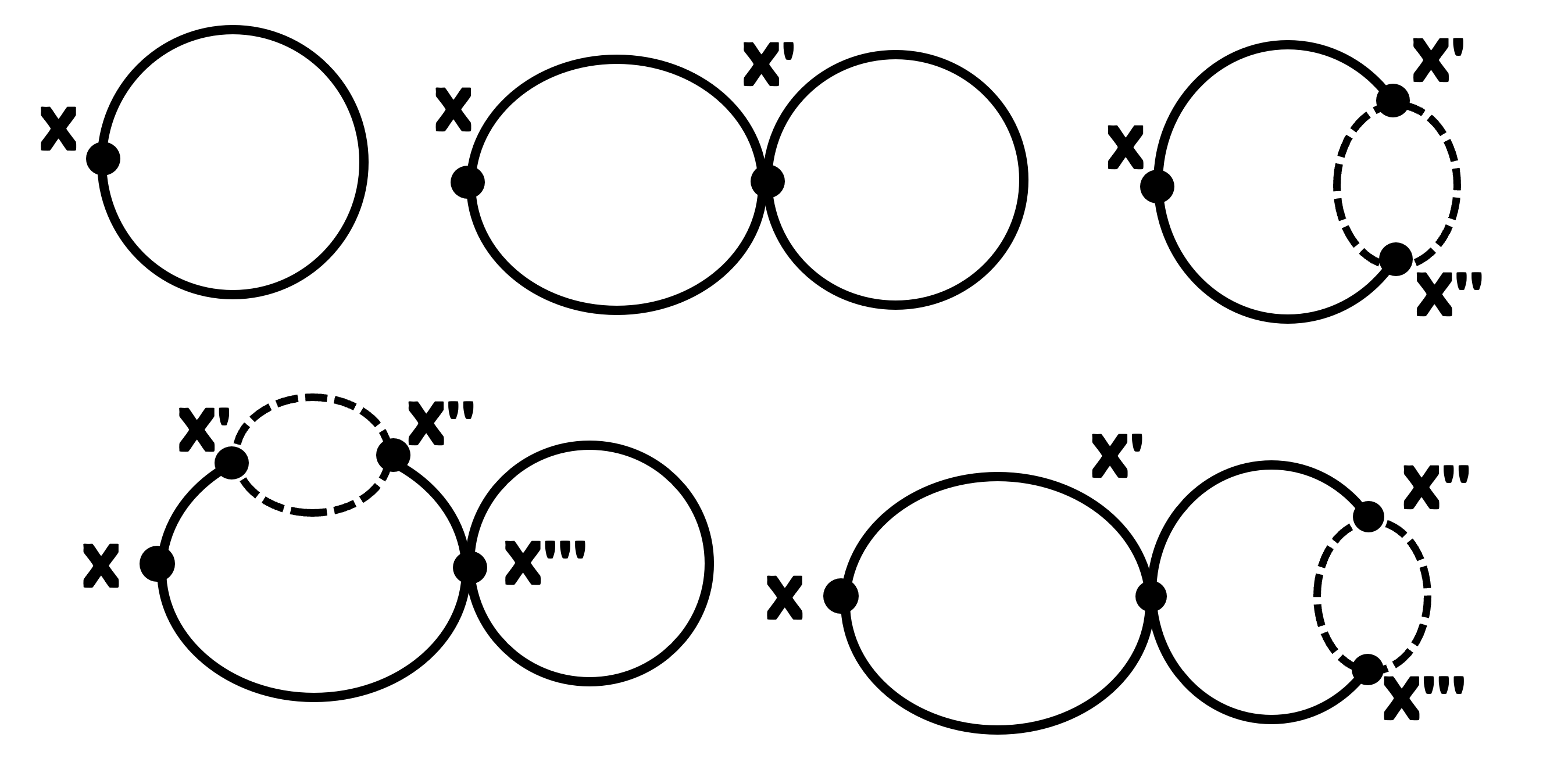}
 \caption{\small \it Feynman diagrams for perturbative computation of $\langle \phi^2(x)\rangle$ up to three loop.  Solid and dashed lines respectively represent the scalar and fermions. All primed vertices  are integrated over.}
  \label{f21}
\end{center}
\end{figure}

\noindent
We shall compute the five loop diagrams for the perturbative $\langle \phi^2 \rangle$, for free theory, ${\cal O}(\lambda)$, ${\cal O}(g^2)$ and ${\cal O}(\lambda g^2)$, as shown in \ref{f21}.  The lowest order or free theory value for $\langle \phi^2 \rangle$ is given by the coincidence limit of any of the four propagators, given by \ref{y6} or \ref{free},
\be
\frac{H^2}{2^2 \pi^2} \ln a
\label{e14-}
\ee

\noindent
At ${\cal O} (\lambda)$, i.e. the second of \ref{f21}, we have 
\begin{eqnarray}
-\frac{i\lambda}{2}  \int d^4 x' a'^4 i\Delta(x',x') \left[i\Delta^2_{++}(x,x')-i\Delta^2_{+-}(x,x') \right]
\label{e14}
\end{eqnarray}
After employing the spatial momentum space, its leading late time secular expression equals
\begin{eqnarray}
-\frac{\lambda H^2}{2^4\times 9\pi^4} \ln^3 a
\label{e15}
\end{eqnarray}

\noindent
At ${\cal O}(g^2)$, i.e. the third of \ref{f21}, we have following the steps described in the preceding section
\begin{eqnarray}
&& -2g^2 {\rm Tr}\int d^4 x' a'^4  \left[i\Delta_{-+}(x,x')i\Delta_{-+}(x,x'')iS^2_{++}(x',x'')+ i\Delta_{+-}(x,x')i\Delta_{+-}(x,x'')iS^2_{--}(x',x'') \right]
\nonumber\\
&&=
-\frac{ig^2}{2^2 \pi^2} \int d^4x' d^4x'' a' a'' \ln (a'a'') (\p^2_{x''} \delta^4(x'-x'') )\left( i\Delta_{+-}(x,x') i\Delta_{+-} (x,x'') - i\Delta_{-+}(x,x') i\Delta_{-+} (x,x'')  \right) \nonumber\\&&=  \frac{g^2 H^2}{2^2\times 3\pi^4} \ln^2 a   
\label{e15a1}
\end{eqnarray}
where we have taken the local part of the fermion loop as earlier.\\ 

\noindent
The fourth of \ref{f21} which is ${\cal O}(\lambda g^2)$, reads
\begin{eqnarray}
&&i\lambda g^2{\rm Tr} \int d^4 x' d^4 x'' d^4 x''' (a'a'' a''')^4 i\Delta (x''',x''')\left[i\Delta_{-+}(x,x') i\Delta_{-+}(x,x''') i\Delta_{++}(x'',x''') i S^2_{++}(x',x'') \right. \nonumber\\&&\left. +i\Delta_{+-}(x,x') i\Delta_{-+}(x,x''') i\Delta_{-+}(x'',x''') i S^2_{--}(x',x'')-i\Delta_{-+}(x,x') i\Delta_{+-}(x,x''') i\Delta_{+-}(x'',x''') i S^2_{++}(x',x'')\right. \nonumber\\&&\left. - i\Delta_{+-}(x,x') i\Delta_{+-}(x,x''') i\Delta_{--}(x'',x''') i S^2_{--}(x',x'')\right]\nonumber\\
&&=\frac{\lambda g^2H^2}{2^5\pi^4} \int d^4 x' d^4 x'' d^4 x''' (a'a'') a'''^4 \ln a''' \ln (a'a'') (\p^2_{x''} \delta^4 (x'-x'')) \left[   i\Delta_{-+}(x,x')i\Delta_{-+}(x,x''')i\Delta_{++}(x'',x''')\right. \nonumber\\&&\left.
- i\Delta_{+-}(x,x')i\Delta_{-+}(x,x''')i\Delta_{-+}(x'',x''')-  i\Delta_{-+}(x,x')i\Delta_{+-}(x,x''')i\Delta_{+-}(x'',x''')+ i\Delta_{+-}(x,x')i\Delta_{+-}(x,x''')i\Delta_{--}(x'',x''') \right] \nonumber\\
\label{e16}
\end{eqnarray}

For $\eta'' \gtrsim \eta'''$, the above becomes
\begin{eqnarray}
&&\frac{\lambda g^2H^2}{2^5\pi^4} \int d^4 x' d^4 x'' d^4 x''' (a'a'') a'''^4 \ln a''' \ln (a'a'') (\p^2_{x''} \delta^4 (x'-x'')) \left(i\Delta_{+-}(x,x') -i\Delta_{-+}(x,x')\right)  \nonumber\\ &&\times \left(i\Delta_{+-}(x,x''')i\Delta_{+-}(x'',x''') -i\Delta_{-+}(x,x''')i\Delta_{-+}(x'',x''')\right) =   -\frac{\lambda g^2 H^2 }{2^7 \times 9\pi^6} \frac{\ln^4 a}{3}
\label{e17}
\end{eqnarray}

For $\eta'' \lesssim \eta'''$ on the other hand, we have
\begin{eqnarray}
&&\frac{\lambda g^2H^2}{2^5\pi^4} \int d^4 x' d^4 x'' d^4 x''' (a'a'') a'''^4 \ln a'''  \ln (a'a'') (\p^2_{x''} \delta^4 (x'-x'') )\left(i\Delta_{+-}(x,x''') -i\Delta_{-+}(x,x''')\right)  \nonumber\\ &&\times \left(i\Delta_{+-}(x,x')i\Delta_{-+}(x'',x''') -i\Delta_{-+}(x,x')i\Delta_{+-}(x'',x''')\right) =   -\frac{\lambda g^2 H^2 }{2^7 \times 9\pi^6} \ln^4 a
\label{e18}
\end{eqnarray}

Combining the above with \ref{e17}, the desired result follows.\\

\noindent
Finally, for the last of \ref{f21}, we have the integral
\begin{eqnarray}
&&i\lambda g^2 {\rm Tr} \int d^4 x' d^4 x'' d^4 x''' (a'a'' a''')^4 \left[ i\Delta_{-+}^2(x,x') i\Delta_{++}(x',x'')i\Delta_{++}(x',x''') iS^2_{++}(x'',x''') \right. \nonumber\\&&\left.+ i\Delta_{-+}^2(x,x') i\Delta_{+-}(x',x'')i\Delta_{+-}(x',x''') iS^2_{--}(x'',x''')- i\Delta_{+-}^2(x,x') i\Delta_{-+}(x',x'')i\Delta_{-+}(x',x''') iS^2_{++}(x'',x''')\right. \nonumber\\&&\left.- i\Delta_{+-}^2(x,x') i\Delta_{--}(x',x'')i\Delta_{--}(x',x''') iS^2_{--}(x'',x''')    \right] \nonumber\\
&&=\frac{\lambda g^2}{2^3\pi^2} \int d^4 x' d^4 x'' d^4 x''' a'^4a'' a''' \ln (a''a''') (\p^2_{x'''}\delta^4(x''-x'''))\left[ i\Delta^2_{-+}(x,x') i\Delta_{++}(x',x'')i\Delta_{++}(x',x''')  \right. \nonumber\\&&\left.- i\Delta^2_{-+}(x,x') i\Delta_{+-}(x',x'')i\Delta_{+-}(x',x''')- i\Delta^2_{+-}(x,x') i\Delta_{-+}(x',x'')i\Delta_{-+}(x',x''')+ i\Delta^2_{+-}(x,x') i\Delta_{--}(x',x'')i\Delta_{--}(x',x''')\right] \nonumber\\
\label{e19}
\end{eqnarray}
The above integral vanishes for $\eta' \lesssim \eta'', \eta'''$. Accordingly it becomes,
\begin{eqnarray}
&&\frac{\lambda g^2}{2^3\pi^2} \int d^4 x' d^4 x'' d^4 x'''  a'^4 (a'' a''') \ln (a'' a''')(\p^2_{x'''}\delta^4(x''-x'''))\nonumber\\&&\times \left( i\Delta^2_{+-}(x,x')- i\Delta^2_{-+}(x,x')\right) \left( i\Delta_{+-}(x',x'')  i\Delta_{+-}(x',x''')- i\Delta_{-+}(x',x'')  i\Delta_{-+}(x',x''')\right) = -\frac{\lambda g^2 H^2 }{2^6 \times 9 \pi^6} \ln^4 a \nonumber\\
\label{e20}
\end{eqnarray}

Combining now \ref{e14-}, \ref{e15}, \ref{e15a1}, \ref{e17}, \ref{e18}, \ref{e20}, we have 
\begin{eqnarray}
\langle \phi^2\rangle =   \frac{H^2}{2^2 \pi^2} \ln a -\frac{\lambda H^2}{2^4\times 9\pi^4} \ln^3 a + \frac{g^2 H^2}{2^2\times 3\pi^4} \ln^2 a- \frac{\lambda g^2 H^2}{2^6\times 9\pi^6} \frac{5\ln^4a}{3}+ {\cal O}(\lambda^2 g^2)
\label{e20add}
\end{eqnarray}
%

%%%%%%
\subsection{Computation of $\langle \overline{\Psi}\Psi \rangle$}
%%%%%%%%%%
%
\begin{figure}[h!]
\begin{center}
  \includegraphics[width=8.0cm]{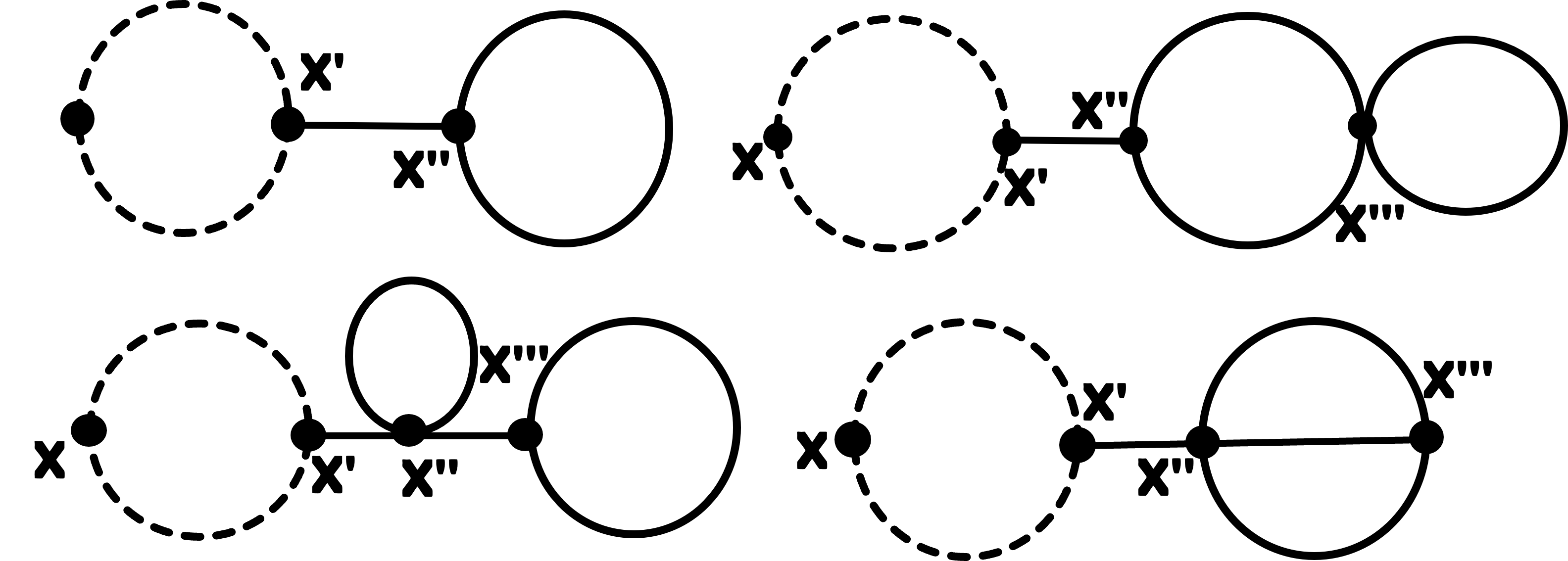}
 \caption{\small \it Loop diagrams for the perturbative computation of $\langle \bar{\Psi}\Psi\rangle$. Dashed and solid lines stand for fermion and scalar respectively. All primed vertices  are integrated over. Also for the free theory, $\langle  \bar{\Psi}\Psi \rangle=0$, as is evident from \ref{y10}.  }
  \label{f22}
\end{center}
\end{figure}

\noindent
There are four diagrams at ${\cal O}(\beta g)$ and ${\cal O}(\beta g \lambda)$ corresponding to $\langle  \bar{\Psi}\Psi (x) \rangle$ up to three loop. For the free theory $\langle  \bar{\Psi}\Psi (x) \rangle=- {\rm Tr}\, iS(x,x)=0$, follows from \ref{y10}. We also note that the fermionic field appearing at the observed vertex at $x$ must be of $+$ kind, in the in-in formalism.  The first of \ref{f22} then gives, recalling that we are interested only in the local contribution from  a closed fermion loop, 
\begin{eqnarray}
&&-\frac{\beta g }{2} \int d^4 x' d^4 x'' (a'a'')^4 {\rm Tr} (i S^2_{++}(x,x'))\left (i\Delta_{++}(x',x'') - i\Delta_{+-}(x',x'')\right) i\Delta (x'',x'')\nonumber\\
&&=-\frac{i\beta g H^2}{2^6\pi^4} \int d^4 x' d^4 x'' \frac{a' a''^4}{a^3}\ln (aa') \ln a'' (\p_{x'}^2\delta^4(x-x')) \left (i\Delta_{+-}(x',x'') - i\Delta_{-+}(x',x'')\right) \nonumber\\&& = \frac{i\beta g H^2}{2^5\pi^4} \int d\eta'' a''^4 \ln a''   \p^2_{\eta}\left(\frac{\ln a}{a^2}  \left (i\Delta_{+-}(0,\eta,\eta'') - i\Delta_{-+}(0,\eta,\eta'')\right) \right) \nonumber\\&&= \frac{i\beta g H}{2^5\pi^4} \int da'' a''^2 \ln a''   \p^2_{\eta}\left(\frac{\ln a}{a^2}\times  \frac{i}{3H a''^3}\right)
= -\frac{\beta gH^2}{2^5 \times 3\pi^4}\ln^3 a
\label{e21}
\end{eqnarray}

\noindent
The second of \ref{f22} reads
\begin{eqnarray}
&&\frac{i\lambda \beta g }{4} \int d^4 x' d^4 x'' d^4 x''' (a'a'' a''')^4 i\Delta(x'',x'')i\Delta(x''',x''')    {\rm Tr} (i S^2_{++}(x,x'))\nonumber\\&&\times   \left (i\Delta_{+-}(x',x'') - i\Delta_{-+}(x',x'')\right)\left (i\Delta_{+-}(x'',x''') - i\Delta_{-+}(x'',x''')\right) \nonumber\\
&&= \frac{\lambda \beta g H^4}{2^9 \times \pi^6}\int d^4 x' d^4 x'' d^4 x''' (a'' a''')^4 \ln a'' \ln a''' \left(\frac{a' \ln (aa')}{a^3} \right) \p^2_{x'}\delta^4(x-x') \nonumber\\
&&\times \left(i\Delta_{+-}(x',x'') - i\Delta_{-+}(x',x'')\right)\left (i\Delta_{+-}(x'',x''') - i\Delta_{-+}(x'',x''')\right)= \frac{\lambda \beta g H^2}{2^{10}\times 9\pi^6} \ln^5 a
\label{e22}
\end{eqnarray}

\noindent
The third diagram of \ref{f22} gives 
\begin{eqnarray}
&&\frac{i\lambda \beta g }{4} \int d^4 x' d^4 x'' d^4 x''' (a'a'' a''')^4 i\Delta(x''',x''')    {\rm Tr} (i S^2_{++}(x,x'))\nonumber\\&&\times   \left (i\Delta_{+-}(x',x'') - i\Delta_{-+}(x',x'')\right)\left (i\Delta^2_{+-}(x'',x''') - i\Delta^2_{-+}(x'',x''')\right) \nonumber\\
&& =  \frac{\lambda \beta g H^2}{2^7 \pi^4} \int d^4 x' d^4 x'' d^4 x''' (a'' a''')^4 \ln a''' \left(\frac{a'}{a^3}\ln (aa') \right) (\p^2_{x'}\delta^4(x-x'))\nonumber\\
&& \times \left (i\Delta_{+-}(x',x'') - i\Delta_{-+}(x',x'')\right)\left (i\Delta^2_{+-}(x'',x''') - i\Delta^2_{-+}(x'',x''')\right) = \frac{\lambda \beta g H^2}{2^8 \times 9\pi^6}\frac{\ln^5 a}{3}
\label{e23}
\end{eqnarray}

\noindent
Finally, the last diagram of \ref{f22} reads
\begin{eqnarray}
&&\frac{i\lambda \beta g }{6} \int d^4 x' d^4 x'' d^4 x''' (a'a'' a''')^4 {\rm Tr} (iS^2_{++}(x,x')) \left (i\Delta_{+-}(x',x'') - i\Delta_{-+}(x',x'')\right)\left (i\Delta^3_{+-}(x'',x''') - i\Delta^3_{-+}(x'',x''')\right) \nonumber\\
&&= \frac{\lambda \beta g H^2}{2^8 \times 9\pi^6} \frac{\ln^5 a}{3}
\label{e24}
\end{eqnarray}

\noindent
Combining now \ref{e14-}, \ref{e15}, \ref{e15a1}, \ref{e17}, \ref{e18}, \ref{e20}, we have 
\begin{eqnarray}
\langle \bar{\psi}\psi\rangle =   -\frac{\beta gH^2}{2^5 \times 3\pi^4}\ln^3 a   + \frac{\lambda \beta g H^2}{2^{10} \times 9 \pi^6} \frac{11\ln^5 a}{3}+   {\cal O}(\lambda^2 \beta g)
\label{e24add}
\end{eqnarray}    
The above result, along with \ref{e13add} and \ref{e20add} completes our perturbative computation associated with $\langle \phi \rangle$.  We next wish to resum this result below in order to find out a non-perturbative expression for the same.

%%%%%%%%%
\subsection{Resummation and non-perturbative $\langle \phi \rangle$}\label{NP1}
%%%%%%
Substituting now \ref{e13add}, \ref{e20add} and  \ref{e24add} into \ref{e2}, we have  for the leading secular logarithms up to the order of the perturbation theory we are interested in, 
\begin{eqnarray}
\frac{d \langle \bar{\phi}\rangle}{d{\cal N}}+  \frac{\bar{\beta}{\cal N} }{2^3\times 3\pi^2} - \left(\frac{\bar{\beta} g^2}{2^5 \times 9\pi^4} + \frac{11\lambda \bar{\beta}}{2^7\times 27\pi^4}\right) {\cal N}^3+ \frac{\lambda \bar{\beta} g^2}{2^{10}\times 9\pi^6} \frac{11{\cal N}^5}{9} =0
\label{e25}
\end{eqnarray}    
where ${\cal N}=Ht$ is the number of $e$-foldings, $\bar{\beta}=\beta/H$ and $\bar \phi=\phi/H$ are dimensionless. Integrating, we have the late time perturbative result
\begin{eqnarray}
\langle \bar{\phi}\rangle =-  \frac{\bar{\beta}{\cal N}^2 }{2^4\times 3\pi^2} + \left(\frac{\bar{\beta} g^2}{2^7 \times 9\pi^4} + \frac{11\lambda \bar{\beta}}{2^9\times 27\pi^4}\right) {\cal N}^4- \frac{\lambda \bar{\beta} g^2}{2^{10}\times 9\pi^6} \frac{11{\cal N}^6}{54} 
\label{e26}
\end{eqnarray}    

The $g=0$ part of the above equation has  also been obtained by directly computing the tadpoles in the IR limit in \ref{C}, for the sake of consistency. They match exactly. This shows that integrating any vacuum diagram considered here basically  corresponds to attaching an external line to the vertex at $x$, and then replacing the original vertex at $x$ by a dummy vertex $x'$ and labelling the new external point on the propagator by $x$, so that  the diagrams of  \ref{fig-C} are generated eventually. Similar argument holds as well when we include the Yukawa interaction. \\

\noindent
Defining now a new variable $\langle \bar\phi \rangle =- \langle \varphi \rangle $ in \ref{e25}, we have 
\begin{eqnarray}
\frac{d \langle {\varphi}\rangle}{d{\cal N}}=  \frac{\bar{\beta}{\cal N} }{2^3\times 3\pi^2} - \left(\frac{\bar{\beta} g^2}{2^5 \times 9\pi^4} + \frac{11\lambda \bar{\beta}}{2^7\times 27\pi^4}\right) {\cal N}^3+ \frac{\lambda \bar{\beta} g^2}{2^{10}\times 9\pi^6} \frac{11{\cal N}^5}{9} 
\label{e27}
\end{eqnarray}    

Following e.g. \cite{Kamenshchik:2020yyn, Kamenshchik:2021tjh},  we wish to integrate the above equation by promoting the $e$-folding ${\cal N}$ to non-perturbative level via inverting \ref{e26} by iteratively using the same into itself.  Keeping in mind the relevant order of the perturbative theory,  we have after some algebra,
\begin{eqnarray}
{\cal N}= \frac{2^2 \sqrt{3}\pi}{\overline{\beta}^{1/2}} \langle \varphi \rangle^{1/2}+ \frac{4\sqrt{3}\pi \langle \varphi \rangle^{3/2}}{\bar{\beta}^{3/2}} \left( g^2+ \frac{11\lambda}{12}\right)+ \frac{\lambda g^2 \pi \langle \varphi \rangle^{5/2}}{\bar{\beta}^{5/2}}\frac{407}{6\sqrt{3}}
\label{e28}
\end{eqnarray}    
which we substitute into \ref{e27} to have, 
\begin{eqnarray}
\frac{d \langle \varphi \rangle }{ d {\cal N}}= \frac{\bar{\beta}^{1/2} \langle \varphi \rangle^{1/2}}{2\sqrt{3}\pi} - \frac{\sqrt{3} \langle \varphi \rangle^{3/2}}{2\pi \bar{\beta}^{1/2}}\left(g^2+\frac{11\lambda}{12} \right) -\frac{649\lambda g^2 }{144\sqrt{3}\pi}\frac{\langle \varphi \rangle^{5/2}}{\bar{\beta}^{3/2}}
\label{e29}
\end{eqnarray}    

Integrating the above equation, we find as ${\cal N}\gg 1$, 
\begin{eqnarray}
\langle \bar \phi \rangle =- \langle \varphi \rangle= \frac{108 \bar{\beta} }{649\lambda g^2} \left( g^2+\frac{11\lambda}{12}\right)\left[1- \left( 1+\frac{649 \lambda g^2}{162 \left(g^2+ \frac{11\lambda}{12}\right)^2} \right)^{1/2} \right]
\label{e30}
\end{eqnarray}    

The above expression is real, vanishes as $\bar{\beta}\to 0$ and flips sign if the cubic coupling does so. On the other hand, we have as $g\to 0$,
\begin{eqnarray}
\langle \bar \phi \rangle_{g\to 0}  =- 0.364\times \frac{\bar\beta}{\lambda} 
\label{e31}
\end{eqnarray}    
The numerical factor $0.364$ appearing above is slightly smaller than the one found in~\cite{Bhattacharya:2022aqi} via direct computation of $\langle \phi \rangle $, using the exact propagators.  This corresponds to a trivial error of overcounting  a factor of $1/3$ there, while computing the third diagram of \ref{fig-C}. (The leading secular expressions for the rest of the diagrams match exactly with that of~\cite{Bhattacharya:2022aqi}). Evidently, there is no such mismatch actually, as has been substantiated by the explicit IR effective computations for the tadpoles in \ref{C}. We also note  from \ref{f0} that the classical minima of the tree level potential is located at $-3\bar{\beta}/\lambda$, approximately one order of magnitude less than \ref{e31}. Thus  these results indeed manifest strong  quantum fluctuations. 

\begin{figure}[h!]
\begin{center}
  \includegraphics[width=6.5cm]{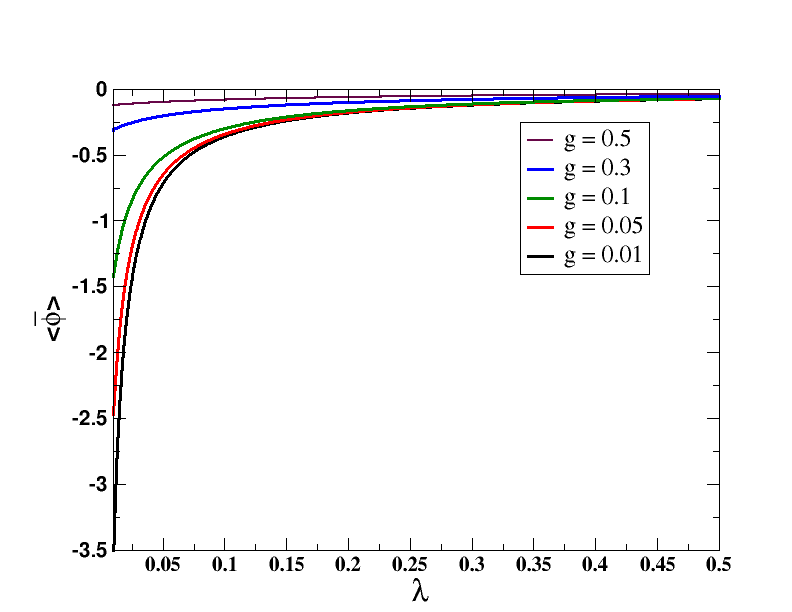}
   \includegraphics[width=6.5cm]{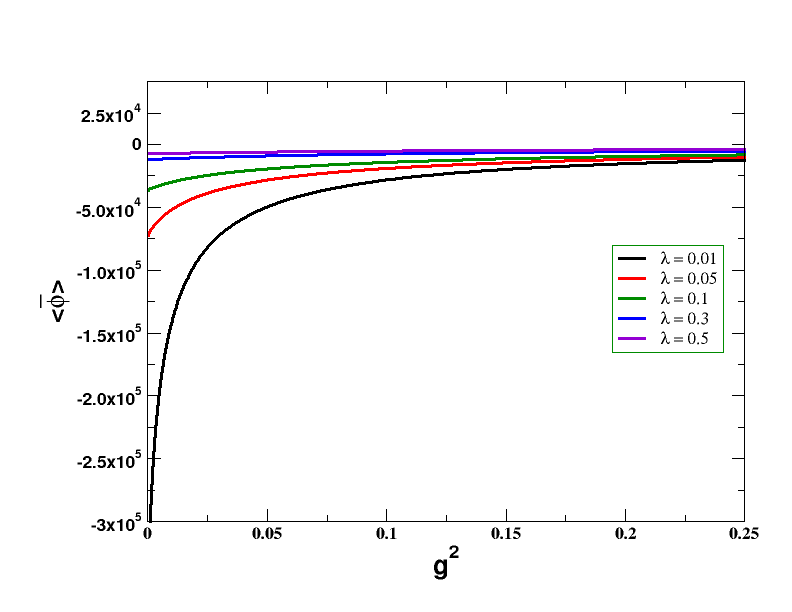}
 \caption{\small \it Variation of the non-perturbative $\langle \bar \phi \rangle$, \ref{e30}, with respect to different couplings. Note that it increases with the increasing cubic coupling as is evident from \ref{e30}, whereas decreases with both increasing quartic and Yukawa couplings. We have  taken $\bar \beta$ to be positive here. For its negative value, $\langle \bar \phi \rangle$ becomes positive. }
  \label{f23}
\end{center}
\end{figure}

\begin{figure}[h!]
\begin{center}
   \includegraphics[width=7.5cm]{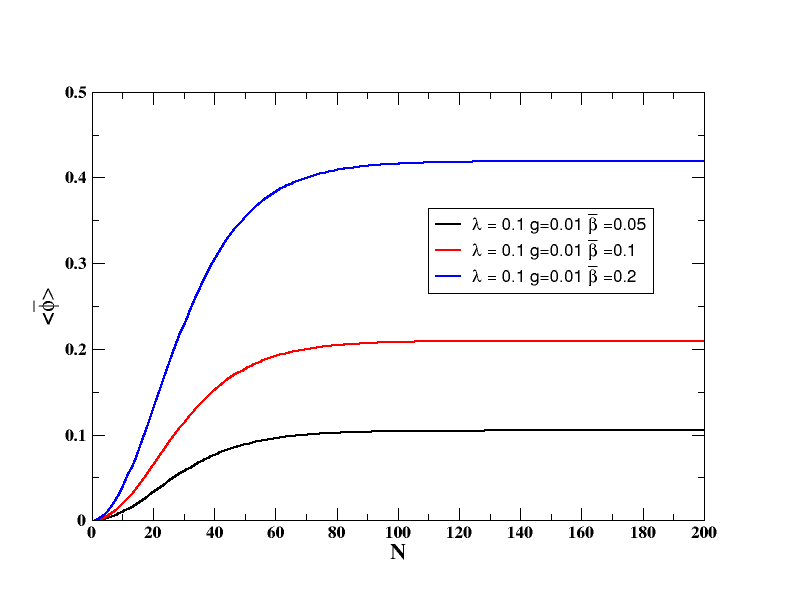}
    \includegraphics[width=7.4cm]{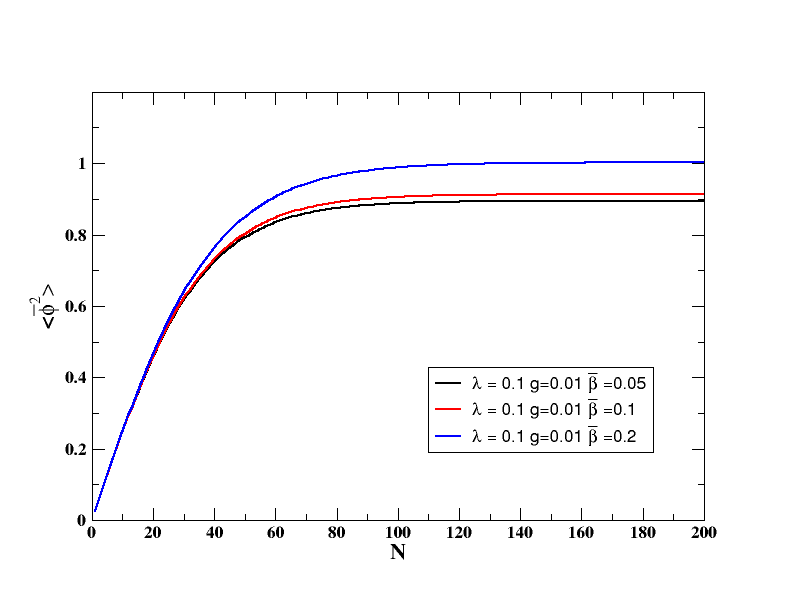}
     \includegraphics[width=7.4cm]{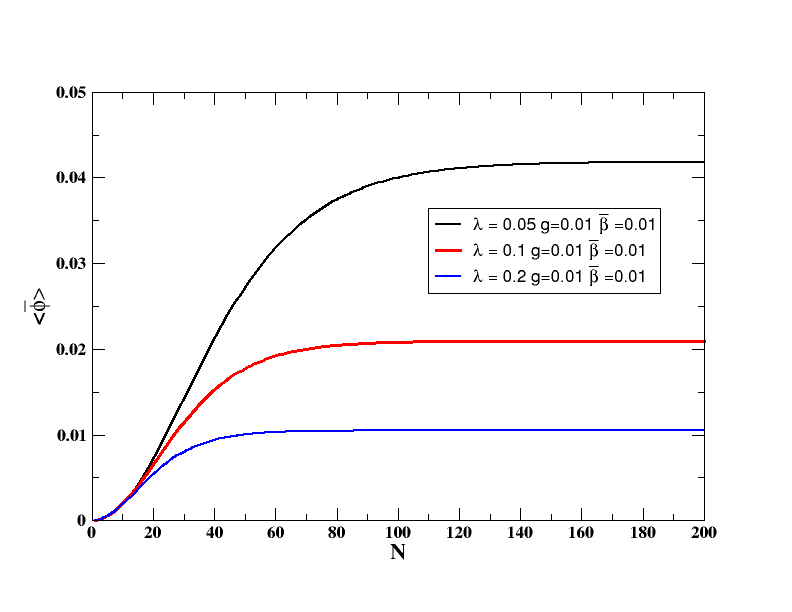}
 \caption{\small \it Variation of $\langle \bar \phi \rangle$ with respect to the number of $e$-foldings, ${\cal N}$, found from \ref{e29} for different values of the couplings.  }
  \label{f24}
\end{center}
\end{figure}

\ref{e30} is one of the main results of this paper. We have plotted the variations of $\langle \bar \phi \rangle$ with respect to different coupling parameters in \ref{f23}. We have also plotted the variation of the same with respect to the number of $e$-foldings in \ref{f24}. \ref{f23} shows that turning on the Yukawa interaction reduces the magnitude of $\langle \bar \phi \rangle$. A non-perturbative $\langle \bar \phi \rangle$ may effectively act as a cosmological constant at late times. We refer our reader to~\cite{Bhattacharya:2022aqi} for an estimation of backreaction due to it. This completes our discussion on the one point function. We now wish to do similar computations for the two point function $\langle \phi^2\rangle$ below.
%
%%%%%%%%%%%
%%%%%%%%%
\section{Perturbative computation of $\langle \phi^2 \rangle$, its resummation and the dynamical mass generation}\label{s4}
%%%%%%
Taking once again the late time and large scale limit of \ref{e1},  multiplying it with the scalar $\phi(x)$, and taking the  expectation value with respect to the initial Bunch-Davies vacuum, we have in the Heisenberg picture 
\begin{eqnarray}
\frac {d \langle \phi^2(x) \rangle} {d {\cal N}} =\frac{H^2}{4\pi^2}- \frac{2}{3H^2} \left[\frac{\lambda}{6} \langle \phi^4(x) \rangle +\frac{\beta}{2}  \langle \phi^3(x)\rangle + g \langle \bar{\psi}\psi \phi (x)\rangle \right]
\label{e33}
\end{eqnarray}    
where ${\cal N}=Ht$ is the number of $e$-foldings as earlier and the first term on the right hand side of the above equation corresponds to the free theory result, \ref{free}. Setting $\bar \beta=0=g$ and employing the Hartree approximation, $\langle \phi^4(x) \rangle= 3\langle \phi^2(x) \rangle^2$, one finds the non-perturbative result, $\langle \phi^2(x) \rangle \sim \lambda^{-1/2}$~\cite{Starobinsky:1994bd}.   This result basically contains resummed superdaisy diagrams for the bubble self energy corresponding to the quartic self interaction. Note that in the Hartree approximation  the function $\langle \phi^2(x) \rangle^2$ is purely local, in the sense that it does not contain any integration over spacetime. Accordingly, it indicates the generation of a dynamical scalar mass of order $\lambda^{1/2}$ at late times, even though the scalar field was massless to begin with.    

We wish to find out a resummed expression for $\langle \phi^2(x) \rangle$ appearing in  \ref{e33}.
The correlators appearing in \ref{e33} will contain contributions from various loops, like self energies. Such loops can be purely local, purely non-local or mixed type. A local contribution  corresponds to the scenario when all the  vertices of a self energy loop shrinks to a single point, by the virtue of a $\delta$-function, often divergent requiring renormalisation. A non-local self energy on the other hand, does not contain any such $\delta$-function, and it contains contribution of  integrals over spacetime. Both self energies show secular divergences, although the non-local part, as we have mentioned in the preceding section,  usually contains greater power of the secular logarithm compared to that of the local part.   We wish to resum below both types of these self energies.  The local part in particular, will give us the dynamically generated  mass of the scalar field at late times. 

%%%%%%%%%
\subsection{ Perturbative computation of $\langle \phi^4\rangle$}
%%%%%%%%%%%
%
\begin{figure}[h!]
\begin{center}
  \includegraphics[width=14.0cm]{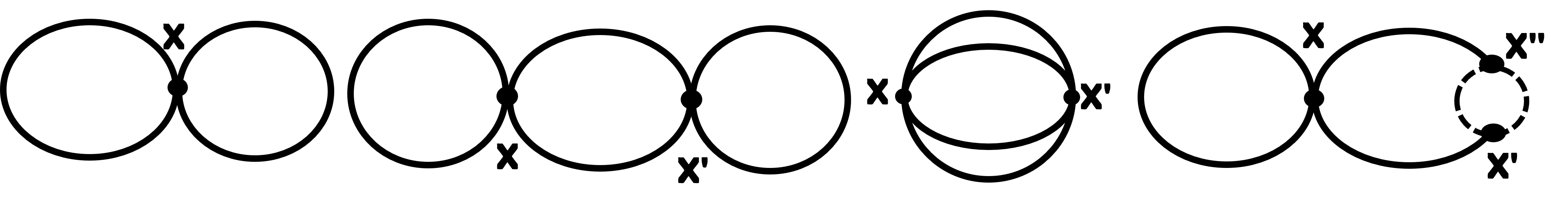}\nonumber\\
   \includegraphics[width=6.0cm]{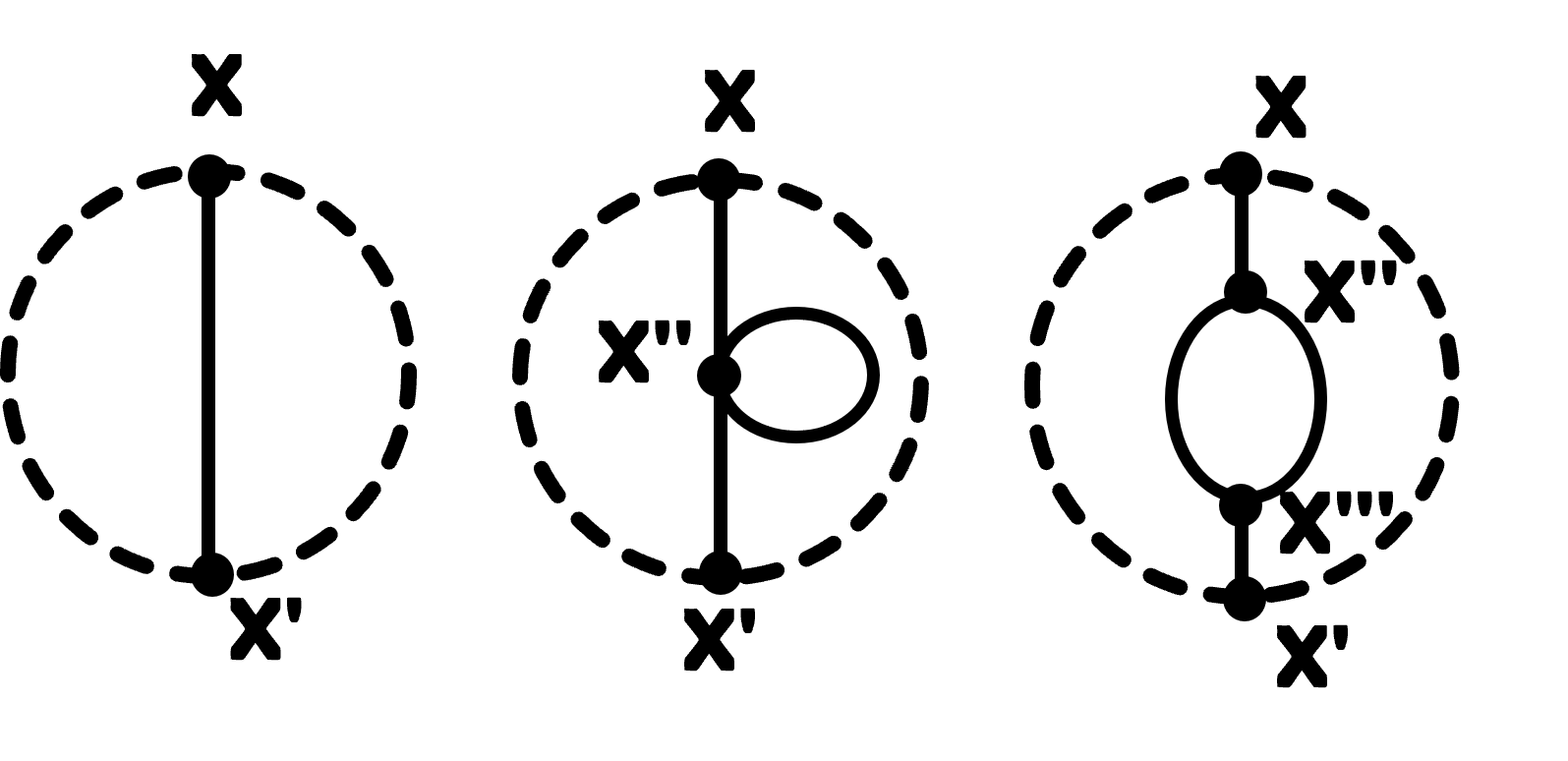}
 \caption{\small \it Vacuum diagrams for \ref{e33} up to three loop. Solid line denotes scalar whereas dashed line represents fermion. Primed vertices are integrated over. We have not included diagrams for $\langle \phi^3 \rangle$ here, as it has already been computed earlier in \ref{SSphi3} (\ref{f2'}). }
  \label{f3'}
\end{center}
\end{figure}
Let us first compute the leading secular contributions in \ref{e33}. Majority of them  are  non-local. We first compute $\langle \phi^4 \rangle$ as given by the first four diagrams of \ref{f3'}. At the lowest order, we have only the local contribution 
\begin{eqnarray}
\langle \phi^4 (x)\rangle = 3 \langle \phi^2 (x)\rangle^2 = \frac{3H^4\ln^2 a }{2^4\pi^4}
\label{e34}
\end{eqnarray}    
where in the last step we have used \ref{free}.\\

\noindent
We now wish to compute  ${\cal O}(\lambda)$ corrections to $\langle \phi^4 (x)\rangle$. The connected bubble diagram (the second of \ref{f3'}) reads 
\begin{eqnarray}
\langle \phi^4 (x)\rangle = -3i\lambda i\Delta(x,x) \times \int d^4 x' a'^4 i\Delta(x',x') \left(i\Delta_{++}^2(x,x')- i\Delta_{+-}^2(x,x') \right)
\label{e35}
\end{eqnarray}    
The leading contribution is non-local and corresponds to the causal part satisfying the temporal hierarchy, $\eta \gtrsim \eta'$. Using the tools of spatial momentum space described in \ref{B}, we find after a little algebra  
\begin{eqnarray}
\langle \phi^4 (x)\rangle = \frac{3i\lambda H^4}{2^4 \pi^4} \ln a \int d^4 x' a'^4 i\ln a'\left(i\Delta_{+-}^2(x,x')- i\Delta_{-+}^2(x,x') \right)= -\frac{\lambda H^4}{2^5 \times 3\pi^6} \ln^4 a
\label{e36}
\end{eqnarray}    

\noindent
The third diagram of \ref{f3'} reads
\begin{eqnarray}
\langle \phi^4 (x)\rangle = -i\lambda  \int d^4 x' a'^4 i\left(i\Delta_{++}^4(x,x')- i\Delta_{+-}^4(x,x') \right)
\label{e37}
\end{eqnarray}    
for which the leading secular part equals
\begin{eqnarray}
&&\langle \phi^4 (x)\rangle = i\lambda  \int d^4 x' a'^4 \left(i\Delta_{+-}^4(x,x')- i\Delta_{-+}^4(x,x') \right)\nonumber\\&&
=i\lambda  \int d^4 x' a'^4 \left(i\Delta_{+-}^2(x,x')+ i\Delta_{-+}^2(x,x') \right) \left(i\Delta_{+-}(x,x')+ i\Delta_{-+}(x,x') \right) \left(i\Delta_{+-}(x,x')- i\Delta_{-+}(x,x') \right)=-\frac{\lambda H^4}{2^6 \times 3\pi^6}\ln^4 a\nonumber\\
\label{e38}
\end{eqnarray}    

\noindent
The fourth  diagram of \ref{f3'} contains correction due to the Yukawa interactions. Recalling that we are only interested in the contribution coming from the local part of the scalar self energy due to the fermion loop, the corresponding integral is written as 
\begin{eqnarray}
\langle \phi^4 (x)\rangle = -12g^2 i\Delta(x,x) {\rm Tr} \int d^4 x' d^4 x'' (a' a'')^4 \left(i\Delta_{++}(x,x')i\Delta_{++}(x,x'') iS^2_{++}(x',x'')  + i\Delta_{+-}(x,x') i\Delta_{+-}(x,x'')iS^2_{--}(x',x'') \right)\nonumber\\
\label{e39}
\end{eqnarray}    

The leading secular contribution for the above integral is given by
\begin{eqnarray}
\frac{g^2 H^4}{2^3 \pi^6}\ln^3 a
\label{e40}
\end{eqnarray}    

Combining now the above with \ref{e34}, \ref{e36}, \ref{e38} we have the perturbative result,
\begin{eqnarray}
\langle \phi^4 (x)\rangle  =  \frac{3H^4\ln^2 a }{2^4\pi^4} - \frac{\lambda H^4}{2^6 \pi^6}\ln^4 a+ \frac{g^2 H^4}{2^3 \pi^6}\ln^3 a  +\dots
\label{e41}
\end{eqnarray}  
%
%%%%%%%%%
\subsection{Perturbative computation of $\langle \phi^3\rangle$}
%%%%%%%%%%%
All the diagrams necessary to compute $\langle \phi^3\rangle$ up to three loop  were considered in \ref{SSphi3}, given by  \ref{f2'} and \ref{e13add}. 

%%%%%%%%%
\subsection{Perturbative computation of $\langle \bar{\psi}\psi \phi\rangle$}
%%%%%%%%%%%

We now wish to evaluate the last three diagrams of \ref{f3'} for computing $\langle \bar{\psi}\psi \phi \rangle$.  The leading order value of $\langle \bar{\psi}\psi \phi\rangle$ comes at ${\cal O}(g)$. Since we are only interested here in the local part of the square of the fermion propagator, it is clear that the diagram corresponding to this leading order (i.e., the third from the last of \ref{f3'}) will only contain purely local contributions,   and  hence we wish to discuss its renormalisation here. Using \ref{y12}, the corresponding integral for $g\langle \bar{\psi}\psi \phi \rangle$ reads
\begin{eqnarray}
&&-ig^2 {\rm Tr} \int a'^d d^d x' i \Delta_{++}(x,x') i S^2_{++}(x,x') = -\frac{\mu^{-\e}g^2\Gamma(1-\e/2)}{2^2 \pi^{2-\e/2} \e (1-\e)}  \int d^dx' \frac{a'^d}{(aa')^{3-\e/2}} i\Delta_{++}(x,x') \p^2_{x'} \delta^d (x-x')\nonumber\\&& -\frac{\mu^{-\e}g^2}{2^3\pi^2} \int d^dx' \frac{a'^d}{(aa')^{3}}\ln (aa') i\Delta_{++}(x,x') \p^2_{x'} \delta^d (x-x')- \frac{3g^2\e}{2^5\pi^2}\int d^4 x' \frac{a'}{a^3}\ln^2 (aa') i \Delta_{++}(x,x') \p^2_{x'}\delta^4(x-x')   \nonumber\\&&
\label{e42}
\end{eqnarray} 

 We now consider the scalar field strength renormalisation counterterm (\ref{y13}) contribution,  by making the replacement
$$ -g^2 {\rm Tr} \,iS^2(x,x')\to \frac{i a^d \delta Z}{ a^2 (aa')^d}$$
This cancels the first term on the right hand side of \ref{e42}, and it gives after integrating by parts
\be 
\frac{\mu^{-\e}g^2 \Gamma(2-\e) H^{4-\e}}{2^{3-\e}\pi^{4-\e/2} \Gamma(1-\e/2)\e} \ln a+ \frac{3g^2 H^4}{2^4\pi^4}\ln^2 a
\label{correctren1}
\ee

In order to renormalise the above contribution, we further add the vacuum expectation value of a conformal counterterm, 
\be
\frac12 \delta \zeta R \langle \phi^2(x)\rangle = \frac{\delta \zeta R H^{2-\e} \Gamma(2-\e)}{2^{3-\e}\pi^{2-\e/2} \Gamma(1-\e/2)} \left(\frac{1}{\e} +\ln a\right)
\label{e42add}
\ee
where we have used \ref{y6}. Hence  the choice 
\be
\delta \zeta R= - \frac{\mu^{-\e} g^2 H^2}{\pi^2 \e} 
\label{e42add1}
\ee
simply cancels the first term  of \ref{correctren1}. The remaining divergence is constant and hence can be absorbed in a cosmological constant counterterm,
\be
\frac{\delta \Lambda}{8\pi G}= \frac{\mu^{-\e} g^2 H^{4-\e} \Gamma(2-\e)}{2^{3-\e} \pi^{4-\e/2} \Gamma(1-\e/2) \e^2}
\label{e42add2}
\ee

Thus the renormalised contribution for \ref{e42} equals 
\be
\frac{3g^2 H^4}{2^4\pi^4}\ln^2 a
\label{e42add3}
\ee

We once again emphasise that the $\delta$-function appearing in \ref{e42} has basically converged all functions to the point $x$,  asserting the term `local'.

\noindent
The sixth and seventh diagrams of \ref{f3'} contain both local and non-local contributions. Let us compute the leading or non-local terms first.   The sixth diagram reads
\begin{eqnarray}
&&-\frac{\lambda g}{2}{\rm Tr} \int d^4 x' d^4 x'' (a'a'')^4 i\Delta(x',x') iS^2_{++}(x,x'')  \left(i\Delta_{++}(x,x')i\Delta_{++}(x'',x')-i\Delta_{+-}(x,x')i\Delta_{+-}(x'',x')   \right)\nonumber\\
&&= - \frac{i\lambda g H^2}{2^5 \pi^4}\int d^4 x' a'^4 \ln a' \p_x^2 \left(\frac{\ln a}{a^2} \left( i\Delta^2_{+-}(x,x')- i\Delta^2_{-+}(x,x')\right)\right)= - \frac{\lambda g H^4}{ 2^5 \times 9\pi^6} \ln^4 a
\label{e43}
\end{eqnarray} 

\noindent
The seventh diagram of \ref{f3'} reads,
\begin{eqnarray}
&& ig \beta^2 {\rm Tr} \int d^4 x' d^4 x'' d^4 x''' (a'a''a''')^4 iS^2_{++}(x,x''')  \left[i\Delta_{++}(x,x')i\Delta_{++}(x''',x'') i\Delta^2_{++}(x',x'') \right. 
\nonumber\\&&\left. - i\Delta_{++}(x,x')i\Delta_{+-}(x''',x'') i\Delta^2_{+-}(x',x'')-i\Delta_{+-}(x,x')i\Delta_{++}(x''',x'') i\Delta^2_{-+}(x',x'')\right. \nonumber\\&&\left. +i\Delta_{+-}(x,x')i\Delta_{+-}(x''',x'') i\Delta^2_{--}(x',x'')      \right]
\label{e44}
\end{eqnarray} 

For $\eta\gtrsim \eta'\gtrsim \eta''$, the above integral becomes 
\begin{eqnarray}
&& \frac{g \beta^2}{2^2\pi^2}  \int d^4 x' d^4 x'' (a'a'')^4 \p_x^2  \left[\frac{\ln a}{a^2}\left(i\Delta_{+-}(x,x')- i\Delta_{-+}(x,x')\right)\left(i\Delta_{+-}(x,x'')i\Delta^2_{+-}(x',x'') -{\rm c.c.}\right) \right]\nonumber\\&&
=\frac{g\beta^2}{2^7\times 9\pi^6}\ln^5 a
\label{e45}
\end{eqnarray} 
whereas for the hierarchy $\eta\gtrsim \eta''\gtrsim \eta'$, the contribution is exactly the same as above. Putting things together now, we have the perturbative result at the leading secular order,
\begin{eqnarray}
\langle \bar{\psi}\psi \phi \rangle = \frac{3gH^4}{2^4\pi^4}\ln^2 a- \frac{\lambda g H^4}{ 2^5 \times 9\pi^6} \ln^4 a+\frac{g\beta^2H^2}{2^6\times 9\pi^6}\ln^5 a+\dots
\label{e46}
\end{eqnarray} 

Substituting  the above along with \ref{e13add}, \ref{e41} into \ref{e33}, we have the perturbative result 
\begin{eqnarray}
 \frac {d\langle \bar{\phi}^2 \rangle} {d{\cal N}} =\frac{1}{4\pi^2}-  \left[\frac{\lambda {\cal N}^2}{2^4\times 3\pi^4} -\frac{\lambda^2 {\cal N}^4}{2^6\times 9\pi^6} -\frac{7\bar{\beta}^2 {\cal N}^3 }{2^6 \times 9\pi^4}+\frac{g^2{\cal N}^2}{2^3 \pi^4}  -\frac{\lambda g^2 {\cal N}^4}{2^4 \times 27\pi^6}+\frac{g^2 \bar{\beta}^2{\cal N}^5}{2^5 \times 27\pi^6} \right]  
\label{e47}
\end{eqnarray} 
so that we  have 
\begin{eqnarray}
 \langle \bar{\phi}^2 \rangle  =\frac{{\cal N}}{4\pi^2}-\frac{\lambda {\cal N}^3}{2^4\times 9\pi^4} +\frac{\lambda^2 {\cal N}^5}{2^6\times 45\pi^6} +\frac{7\bar{\beta}^2 {\cal N}^4 }{2^6 \times 36\pi^4}-\frac{g^2{\cal N}^3}{2^3\times 3\pi^4}  +\frac{\lambda g^2 {\cal N}^5}{2^4 \times 135\pi^6}-\frac{g^2 \bar{\beta}^2{\cal N}^6}{2^5 \times 162\pi^6}  
\label{e48}
\end{eqnarray} 
where $\bar{\phi}=\phi/H$ and $\bar{\beta}=\beta/H$ are dimensionless as before. Promoting now \ref{e47} to non-perturbative level as earlier, we have 
\begin{eqnarray}
\frac {d\langle \bar{\phi}^2 \rangle} {d{\cal N}} =\frac{1}{4\pi^2}-\frac{\lambda \langle \bar{\phi}^2 \rangle^2}{3} +\frac{4\pi^2\lambda^2\langle \bar{\phi}^2\rangle^4}{27}  -   \frac{80\pi^2\lambda g^2 \langle \bar{\phi}^2\rangle^4}{27}     +\frac{7 \pi^2 \bar{\beta}^2 \langle \bar{\phi}^2\rangle^3 }{9}- 2g^2 \langle \bar{\phi}^2\rangle^2    +\frac{52 \pi^4 g^2 \bar{\beta}^2 \langle \bar{\phi}^2\rangle^5 }{ 27}  
\label{e49}
\end{eqnarray} 

The solution to the above equation turns out to be complex for certain values in the parameter space. This is unacceptable, since $\phi(x)$ is hermitian and hence $\langle \phi^2 \rangle$ must be a real, positive definite quantity.  It can be readily verified  that  the above equation can be integrated only if we ignore the cubic coupling term. In that case we have for large ${\cal N}$,
\begin{eqnarray*}
 \langle \bar{\phi}^2 \rangle =  \frac{3 \sqrt{6g^2+\lambda}}{4\pi g \sqrt{10\lambda} }\left[-1+ \left(1 +\frac{80\lambda g^2}{3(6g^2+\lambda)^2}\right)^{1/2} \right]^{\frac12} 
\label{e50}
\end{eqnarray*} 

However, if we understand it correctly, the above result seems to be a bit misleading! This is because  we have not attempted to distinguish between the purely local and non-local terms in \ref{e49}. Let us now consider the purely non-local terms  
\begin{eqnarray}
\frac {d\langle \bar{\phi}^2 \rangle_{\rm NL}} {d{\cal N}} =\frac{4\pi^2\lambda^2\langle \bar{\phi}^2\rangle^4_{\rm NL}}{27}  -   \frac{80\pi^2\lambda g^2 \langle \bar{\phi}^2\rangle^4_{\rm NL}}{27}     +\frac{7 \pi^2 \bar{\beta}^2 \langle \bar{\phi}^2\rangle^3_{\rm NL} }{9}  +\frac{52 \pi^4 g^2 \bar{\beta}^2 \langle \bar{\phi}^2\rangle^5_{\rm NL} }{ 27}  
\label{e51'}
\end{eqnarray} 
which gives
\begin{eqnarray}
\int \frac{d\langle \bar{\phi}^2 \rangle_{\rm NL}}{ \langle \bar{\phi}^2\rangle^3_{\rm NL}\left[ 7\bar{\beta}^2 + \frac{4\lambda}{3}\left(\lambda- 20 g^2 \right)\langle \bar{\phi}^2 \rangle_{\rm NL}  +\frac{52 \pi^2 g^2 \bar{\beta}^2}{3}\langle \bar{\phi}^2 \rangle_{\rm NL}^2 \right]    } =\frac{\pi^2 {\cal N}}{9}
\label{e52'}
\end{eqnarray} 
It is easy to check that the quadratic cannot have real and positive roots for all values of the couplings. Could it possibly indicate that $\langle\bar{\phi}^2 \rangle_{\rm NL}$ is vanishing? Even though we shall not investigate the asymptotic behaviour of $\langle\bar{\phi}^2 \rangle_{\rm NL}$ explicitly in this paper, we wish to comment a little bit more on this issue  towards the end of the following section, and will argue that possibly a  more careful analysis is necessary to predict anything specific about the same. We also note in this context that for a quartic self interaction, it was argued in~\cite{Youssef:2013by} using the Schwinger-Dyson equation for the Feynman propagator that the same can admit one asymptotically vanishing non-perturbative solution. See also the discussion of~\cite{Kamenshchik:2020yyn} on  resummed two point correlation function for quartic self interaction beyond local approximation.

%%%%%%%%
\subsection{The purely local contribution and the dynamical mass}\label{loc-comp}
%%%%%%%%%%%%

Let us now consider the purely local contributions to $\langle \phi^2(x)\rangle$. The renormalised $\langle \phi^2(x)\rangle $ containing contributions only from the local part of the self energy at ${\cal O}(\lambda)$, ${\cal O}(\lambda^2)$ and ${\cal O}(\beta^2)$ for one particle irreducible (1PI)  diagrams are given by \cite{Bhattacharya:2022wjl}, 
\begin{eqnarray}
\langle \phi^2(x) \rangle_{\rm loc., Ren.} = -\frac{\lambda H^2 {\cal N}^3}{2^4\times 9\pi^4} + \frac{\beta^2 {\cal N}^3}{2^5\times 9\pi^4}+\frac{\lambda^2 H^2 {\cal N}^4}{2^9 \times 3 \pi^6}
\label{e51}
\end{eqnarray}    

\noindent
Let us now consider the local contribution due to ${\cal O}(g^2)$ loop correction to the $\lambda \phi^4$ term in \ref{e33}. It reads
\begin{eqnarray}
&&\frac{i\lambda g^2}{2^2} i\Delta(x,x) {\rm Tr} \int d^d x' d^d x'' (a'a'')^d iS^2_{++}(x',x'') i\Delta_{++}(x,x')i\Delta_{++}(x,x'')\nonumber\\
&&= \frac{\mu^{-\e}\lambda g^2 \Gamma(1-\e/2)}{2^4 \pi^{2-\e/2}\e(1-\e)} i\Delta(x,x)\int d^d x' d^d x'' \frac{(a'a'')^d}{(a'a'')^{3-\e}} (\p_{x''}^2 \delta^d(x'-x'') )i\Delta_{++}(x,x')i\Delta_{++}(x,x'')\nonumber\\
&&=\frac{\mu^{-\e}\lambda g^2 \Gamma(1-\e/2)}{2^4 \pi^{2-\e/2}\e(1-\e)} i\Delta(x,x)\int d^d x' \p_{x'}^2\left( a'^2 i\Delta^2_{++}(x,x')\right)
\label{e51a1}
\end{eqnarray}    
where we have used \ref{y12}.  Now, the local part of $i\Delta^2_{++}(x,x')$  contains a $\delta$-function,~\ref{e51'}. The integral $\p_{x'}^2\left( a'^2 i\Delta^2_{++}(x,x')\right)$ can be converted to a boundary integral making it vanishing. Similarly the ${\cal O}(\beta g^2)$ local correction to the cubic potential term also vanishes. This happens due to the fact that the scalar self energy due to fermion loop  is accompanied by a $\p^2$. However, there is certainly non-vanishing non-local contributions, as evaluated in \ref{e39}, \ref{e40}.  \\

\noindent
Let us now consider the Yukawa term $g \langle \bar{\psi}\psi \phi \rangle$ in \ref{e33}. Its leading local and renormalised expression was computed in the preceding section, given by~\ref{e42add3}.\\

\noindent
We  now wish to compute the local contributions  at ${\cal O}(\lambda g^2)$ and ${\cal O}(\beta^2 g^2)$ for the same, where one Yukawa interaction term sits at the unintegrated vertex, $x$ (the last two diagrams of \ref{f3'}). We shall see that these local contributions are non-vanishing unlike the above.  We have at ${\cal O}(\lambda g^2)$,
\begin{eqnarray}
&&-\frac{\lambda g^2}{2} {\rm Tr} \int d^d x' d^d x'' (a'a'')^d iS^2_{++}(x,x') i\Delta_{++}(x,x'') i\Delta_{++}(x',x'') i\Delta(x'',x'') \nonumber\\&&
=-\frac{\lambda g^2 H^{2-\e} \Gamma(2-\e)}{2^{3-\e} \pi^{2-\e/2} \Gamma(1-\e/2)} {\rm Tr} \int d^d x' d^d x'' (a'a'')^d iS^2_{++}(x,x') i\Delta_{++}(x,x'') i\Delta_{++}(x',x'')\left(\frac{1}{\e}+\ln a'' \right) 
\label{e52}
\end{eqnarray}    

We add to the above the contribution coming from the one loop mass renormalisation counterterm due to quartic self interaction, $\delta m^2_{\lambda}$ given by  \ref{y7},
\begin{eqnarray}
&&-g^2 \delta m_{\lambda}^2 {\rm Tr}\int (a'a'')^d d^dx' d^d x'' i S_{++}^2(x,x') i\Delta_{++}(x,x'')i\Delta_{++}(x',x'')\nonumber\\&& = \frac{\lambda H^{2-\e}\Gamma(2-\e)}{2^{3-\e} \pi^{2-\e/2} \Gamma(1-\e/2)\e}{\rm Tr}\int (a'a'')^d d^dx' d^d x'' i S_{++}^2(x,x') i\Delta_{++}(x,x'')i\Delta_{++}(x',x'')
\label{e53}
\end{eqnarray}    
so that after using \ref{y12} for the square of the fermion propagator, the above equation becomes
\begin{eqnarray}
&&\frac{i\mu^{-\e}\lambda g^2 H^{2-\e} \Gamma(1-\e)}{2^{5-\e} \pi^{4-\e} \e }  \int d^d x' d^d x'' \frac{(a'a'')^d}{(aa')^{3-\e/2}} (\p^2_{x'} \delta^d(x-x')) i\Delta_{++}(x,x'') i\Delta_{++}(x',x'')\ln a''  \nonumber\\&&
+\frac{i\mu^{-\e}\lambda g^2 H^{2-\e} \Gamma(2-\e)}{2^{6-\e} \pi^{4-\e/2} \Gamma (1-\e/2)}  \int d^d x' d^d x'' \frac{(a'a'')^d}{(aa')^{3}}\ln (aa') (\p^2_{x'} \delta^d(x-x')) i\Delta_{++}(x,x'') i\Delta_{++}(x',x'')\ln a'' \nonumber\\&&
+\frac{3i\e \mu^{-\e}\lambda g^2 H^{2-\e} \Gamma(2-\e)}{2^{8-\e} \pi^{4-\e/2} \Gamma (1-\e/2)}  \int d^d x' d^d x'' \frac{(a'a'')^d}{(aa')^{3}}\ln^2 (aa') (\p^2_{x'} \delta^d(x-x')) i\Delta_{++}(x,x'') i\Delta_{++}(x',x'')\ln a'' 
\label{e54}
\end{eqnarray}    

Let us consider the first integral above. It can as earlier be canceled by the scalar field strength renormalisation counterterm,  \ref{y13}. The second integral turns out to be  completely ultraviolet divergent 
\begin{eqnarray}
-\frac{\mu^{-2\e}\lambda g^2 H^{4-\e} \Gamma(1-\e) }{2^{7-\e} \pi^{6-\e}} \frac{\ln^2 a}{\e}
\label{e56}
\end{eqnarray}    
whereas the third integral pf \ref{e54} is completely ultraviolet finite,
\begin{eqnarray}
- \frac{3\lambda g^2 H^4}{2^8 \pi^6}\ln^3 a
\label{e57}
\end{eqnarray}    

Thus we identify the ultraviolet divergent and finite  parts of \ref{e54} as
\begin{eqnarray}
- \frac{\mu^{-2\e}\lambda g^2 H^{4-\e} \Gamma(1-\e) }{2^{7-\e} \pi^{6-\e}} \frac{\ln^2 a}{\e}-\frac{3\lambda g^2 H^4}{2^8 \pi^6}\ln^3 a
\label{e57}
\end{eqnarray}    

In order to renormalise the above expression, we add with it the ${\cal O}(\lambda)$ corrections to the conformal counterterm introduced in \ref{e42add1}. These corrections respectively correspond to the quartic self interaction and the one loop  mass counterterm corresponding to the quartic self interaction. They read
\begin{eqnarray}
&&-\frac{i\lambda \delta \zeta R}{2^2} \int d^d x' a'^d i\Delta_{++}^2 (x,x') i\Delta(x',x') -\frac{i\delta \zeta R\, \delta m_{\lambda}^2}{2} \int d^d x' a'^d i \Delta^2_{++}(x,x')\nonumber\\&&
=\frac{\mu^{-2\e} \lambda g^2 H^{4-\e} \Gamma(1-\e)}{2^{7-\e}\pi^{6-\e}} \left(\frac{1}{\e^3}+\frac{2\ln a}{\e^2}+\frac{3\ln^2 a}{2\e} +\frac23 \ln^3 a \right)
-\frac{\mu^{-2\e} \lambda g^2 H^{4-\e} \Gamma(1-\e)}{2^{7-\e}\pi^{6-\e}} \left(\frac{1}{\e^3}+\frac{\ln a}{\e^2}+\frac{\ln^2 a}{2\e} +\frac16 \ln^3 a \right)\nonumber\\&&
= \frac{\mu^{-2\e} \lambda g^2 H^{4-\e} \Gamma(1-\e)}{2^{7-\e}\pi^{6-\e}}\left( \frac{\ln a}{\e^2}+\frac{\ln^2 a}{\e}\right) +\frac{\lambda g^2 H^4}{2^7 \pi^6} \ln^3 a
\label{e58}
\end{eqnarray}    
The above expression, when combined with \ref{e57}, cancels the  ultraviolet divergence for the $\ln^2 a$ term. The remaining logarithmic divergence can be canceled as earlier by a conformal counterterm of ${\cal O}(\lambda g^2)$, leaving us with the renormalised expression for the integral
\ref{e52},
\be
-\frac{\lambda g^2 H^4}{2^8 \pi^6} \ln^3 a
\label{e59}
\ee

The diagram at ${\cal O}(\beta^2 g^2)$ can be computed in a similar manner. It reads 
\begin{eqnarray}
&&\frac{i\beta^2 g^2}{2}{\rm Tr} \int d^d x' d^d x'' d^d x''' (a'a'' a''')^d i S^2_{++}(x,x') i\Delta_{++}(x,x'')i\Delta_{++}(x',x''')i\Delta^2_{++}(x'',x''')\nonumber\\
&&=\frac{\mu^{-\e}\beta^2 g^2 \Gamma(1-\e/2)}{2^4 \pi^{2-\e/2}\e(1-\e)} {\rm Tr} \int d^d x' d^d x''  (a'a'')^d i S^2_{++}(x,x') i\Delta_{++}(x,x'')i\Delta_{++}(x',x'')\left(1+\e \ln a'' + \frac{\e^2}{2}\ln^2 a'' + {\cal O}(\e^3) \right)\nonumber\\
%= \frac{\mu^{-3\e}\beta^2 g^2 H^2 \Gamma^3(1-\e/2)}{2^8\pi^{6-3\e/2} (1-\e)^3}\left(\frac{1}{\e^3}+\frac{3\ln a}{\e^2}+\frac{9\ln^2 a}{2\e}+\frac{9}{2}\ln^3 a \right)
\label{e60}
\end{eqnarray}    
We add with it the one loop mass renormalisation counterterm contribution due to the cubic self interaction, \ref{e51'add}, to have
\begin{eqnarray}
&&-\delta m_{\beta}^2 g^2 {\rm Tr} \int d^d x' d^d x''  (a'a'')^d i S^2_{++}(x,x') i\Delta_{++}(x,x'')i\Delta_{++}(x',x'')
%&&= -\frac{\mu^{-3\e}\beta^2 g^2 H^2 \Gamma^3(1-\e/2)}{2^8\pi^{6-3\e/2} (1-\e)^3}\left(\frac{1}{\e^3}+\frac{2\ln a}{\e^2}+\frac{2\ln^2 a}{\e}+\frac{4}{3}\ln^3 a \right)
\label{e61}
\end{eqnarray}    
which cancels the first divergence (i.e., the first term) of \ref{e60}. We now have 
\begin{eqnarray}
&&-\frac{i\mu^{-2\e}\beta^2 g^2 \Gamma^2(1-\e/2)}{2^6 \pi^{4-\e} \e (1-\e)^2} \int d^d x'' \p_x^2\left(\frac{(aa'')^d}{a^{6-\e}}i\Delta^2_{++}(x,x'') f(a'')\right)\nonumber\\
&&-\frac{i\mu^{-2\e}\beta^2 g^2 \Gamma(1-\e/2)}{2^6 \pi^{4-\e/2} (1-\e)} \int d^d x'' \p_x^2\left(\frac{(aa'')^d}{a^{6}}\ln a\, i\Delta^2_{++}(x,x'') f(a'')\right)\nonumber\\
&&-\frac{3i\e \mu^{-2\e}\beta^2 g^2 \Gamma(1-\e/2)}{2^7 \pi^{4-\e/2} (1-\e)} \int d^d x'' \p_x^2\left(\frac{(aa'')^d}{a^{6}}  \ln^2 a\,i\Delta^2_{++}(x,x'') f(a'')\right)
%= \frac{\mu^{-3\e}\beta^2 g^2 H^2 \Gamma^3(1-\e/2)}{2^8\pi^{6-3\e/2} (1-\e)^3}\left(\frac{1}{\e^3}+\frac{3\ln a}{\e^2}+\frac{9\ln^2 a}{2\e}+\frac{9}{2}\ln^3 a \right)
%&&=
\label{e62}
\end{eqnarray}    
where we have abbreviated for the sake of convenience
$$f(a)=\ln a + \frac{\e}{2}\ln^2 a$$
The first line of the above equation can as earlier be canceled by the scalar field strength renormalisation counterterm's contribution. The remaining integrals give
\be
\frac{\mu^{-3\e}\beta^2 g^2 H^2 \Gamma^2(1-\e/2)}{2^8 \pi^{6-\e} (1-\e)^2}\left(\frac{\ln^2 a}{\e} +\frac12\ln^3 a \right)+\frac{3\beta^2 H^2 g^2 }{2^9 \pi^6}\ln^3 a
\label{e62a1}
\ee

We consider the ${\cal O}(\beta^2)$ correction to the conformal counterterm, \ref{e42add1},
\begin{eqnarray}
&&-\frac{\delta \zeta \beta^2 R}{2^2} \int d^d x' d^d x'' (a'a'')^d i\Delta_{++}(x,x')i\Delta_{++}(x,x'')i\Delta^2_{++}(x',x'')-\frac{\delta \zeta R \delta m^2_{\beta}}{2} \int d^d x' a'^d i\Delta^2_{++}(x,x') \nonumber\\&&
=-\frac{\mu^{-3\e} \beta^2 g^2 H^2 \Gamma^2(1-\e/2)}{2^8\pi^{6-\e} (1-\e)^2}\left(\frac{1}{\e^3}+\frac{2\ln a}{\e^2}+\frac{2\ln^2 a}{\e}+\frac43 \ln^3 a \right)+\frac{\mu^{-3\e} \beta^2 g^2 H^2 \Gamma^2(1-\e/2)}{2^8\pi^{6-\e} (1-\e)^2}\left(\frac{1}{\e^3}+\frac{\ln a}{\e^2}+\frac{\ln^2 a}{2\e}+\frac16 \ln^3 a \right)\nonumber\\
\label{e63}
\end{eqnarray}    

The above, when combined with \ref{e62a1}, yields
\begin{eqnarray}
\frac{\mu^{-3\e} \beta^2 g^2 H^2 \Gamma^2(1-\e/2)}{2^8\pi^{6-\e} (1-\e)^2}\left(-\frac{\ln a}{\e^2}-\frac{\ln^2 a}{2\e} \right)+  \frac{5\beta^2 H^2 g^2 }{2^9\times 3\pi^6}\ln^3 a
\label{e64}
\end{eqnarray}    

Now, the logarithmic divergence appearing above can be absorbed by an ${\cal O}(\beta^2 g^2)$ conformal counterterm as earlier. However unlike the quartic self interaction, the $\ln^2 a$ remains. Probably we need to consider terms where both local and non-local terms coincide  and the divergence due to the local term cancels the same. We shall not further pursue it here and instead proceed with the finite term appearing in \ref{e64}. Combining this with \ref{e42add3}, \ref{e51} and  \ref{e59}, we have from \ref{e33} for the local contributions
\begin{eqnarray}
\frac{d\langle \bar{\phi}^2 \rangle_{\rm loc}}{d{\cal N}}= \frac{1}{4\pi^2} -\left(g^2+\frac{\lambda}{6}-\frac{\bar{\beta}^2}{12} \right)\frac{{\cal N}^2}{2^3\pi^4}+ \left(\lambda^2+\lambda g^2 -\frac{5\bar{\beta}^2 g^2}{6} \right)\frac{{\cal N}^3}{2^7\times 3\pi^6}
\label{e65}
\end{eqnarray}    
which reads when promoted to non-perturbative level
\begin{eqnarray}
\frac{d\langle \bar{\phi}^2 \rangle_{\rm loc}}{d{\cal N}}= \frac{1}{4\pi^2} -2\left(g^2+\frac{\lambda}{6}-\frac{\bar{\beta}^2}{12}\right) \langle \bar{\phi}^2 \rangle^2_{\rm loc}+\frac16 \left(\lambda^2+\lambda g^2-\frac{5\bar{\beta}^2g^2}{6}\right)\langle \bar{\phi}^2 \rangle^3_{\rm loc}
\label{e66}
\end{eqnarray}    
However, the above integration gives complex $\langle \bar{\phi}^2 \rangle_{\rm loc} $ for certain parameter values, which is once again unacceptable. We can encounter this issue as follows. Note that here we are dealing with the diagrams of \ref{f3'} in order to compute the first derivative of  $\langle \bar{\phi}^2 \rangle $. Naturally, as we have also discussed below \ref{e26} in the context of tadpoles, while integrating this expression we basically introduce an external point on each of these  diagrams where two scalar propagators meet. Consequently, the original unintegrated vertex at $x$ now becomes a dummy vertex. This gives us the vacuum expectation value of the scalar two point function in the coincidence limit. 

\begin{figure}[h!]
\begin{center}
  \includegraphics[width=10.0cm]{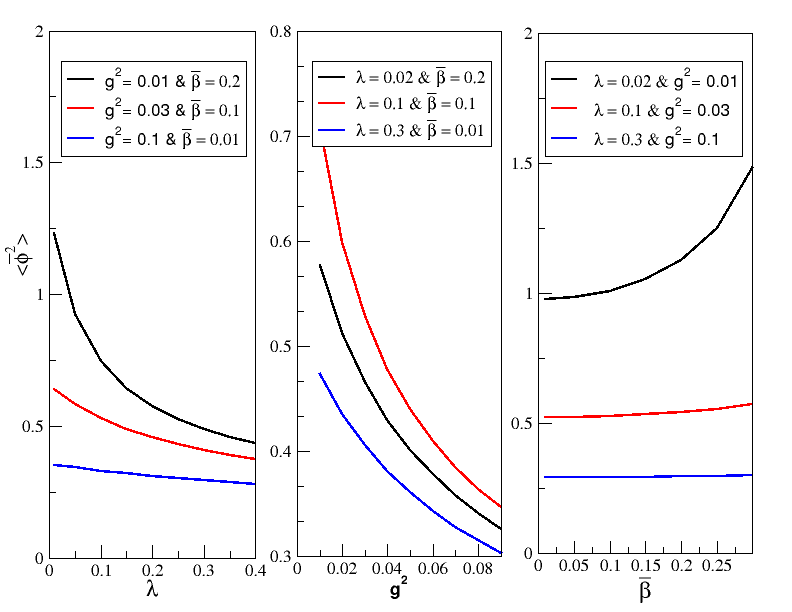}\nonumber\\
  \caption{\small \it The variation of $\langle \bar{\phi}^2 \rangle_{\rm loc}$ with respect to different couplings. See main text for discussion.  }
  \label{f4}
\end{center}
\end{figure}

\begin{figure}[h!]
\begin{center}
   \includegraphics[width=7cm]{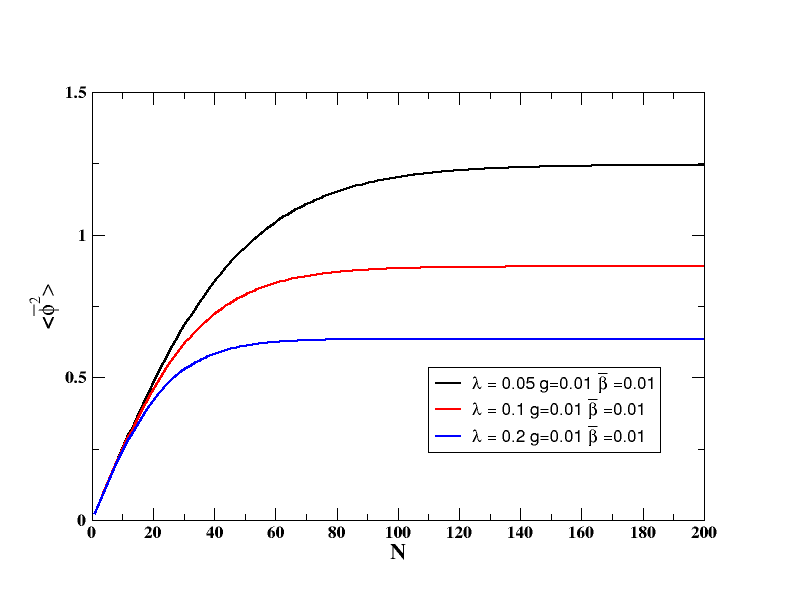}
    \includegraphics[width=7cm]{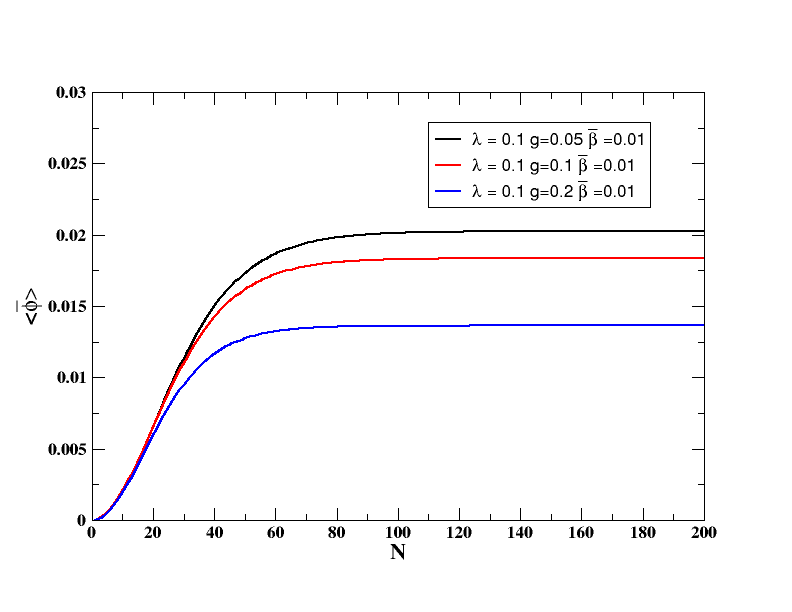}
     \includegraphics[width=7.5cm]{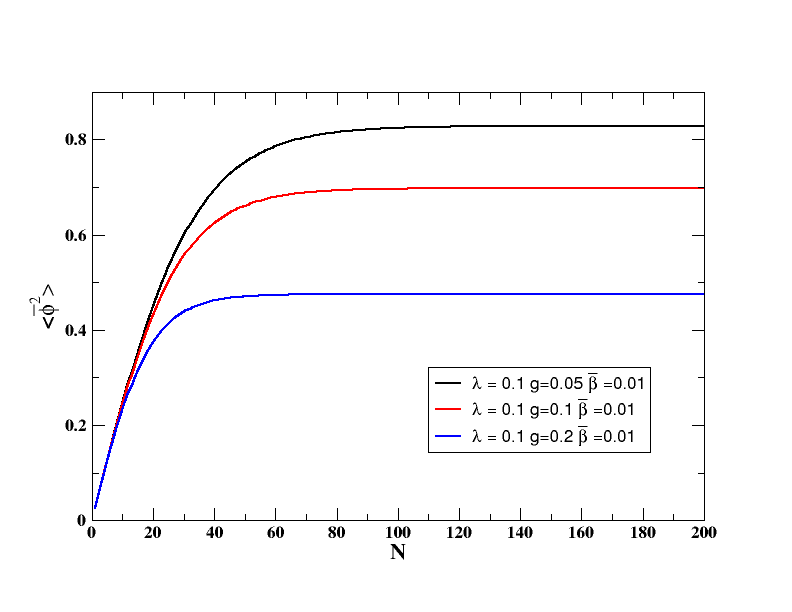}
 \caption{\small \it The variation of $\langle \bar{\phi}^2 \rangle_{\rm loc}$ with respect to the number of $e$-foldings, ${\cal N}$. See main text for discussion. }
  \label{f5}
\end{center}
\end{figure}

As has been done in \ref{C} for $\langle \phi \rangle$, let us now perform an explicit computation for $\langle \phi^2 \rangle$ by considering e.g. the first of \ref{f3'}. When its contribution is integrated with respect to the $e$-foldings, we obtain the second term on the right hand side of  \ref{e48}. As we have stated above, at the diagrammatic level for $\langle \phi^2 \rangle$, this should be equivalent to introducing an external point $x$ on any of the loops of the first of \ref{f3'}, and subsequently to replace the original vertex at $x$ by a dummy one, say $x'$, giving us the ${\cal O}(\lambda)$  contribution,
$$ \langle \phi^2(x) \rangle = -\frac{i\lambda}{2} \int d^4 x a^4 i\Delta(x',x') \left[ i\Delta^2_{++}(x,x')-i\Delta^2_{+-}(x,x')\right] $$ 
It can be easily checked that (e.g.~\cite{Bhattacharya:2022wjl}) the above exactly matches with the aforementioned result of \ref{e48}.
Now, since these diagrams for the two point correlator will contain self energy loops, we should take the  1PI diagrams only. However, it is easy to see that there is no such 1PI diagrams corresponding to the last to diagrams of \ref{f3'}. This is because the corresponding integrated version of these two diagrams will contain the external point between $x''$ and the initial unintegrated vertex at $x$. Hence it seems reasonable to discard them.  Note that we do not need to worry about similar things while computing the scalar one point function, for any diagram corresponding to $\langle \phi \rangle$ contains tadpole (connected to an external point) which cannot be  1PI.  Indeed, after ignoring the ${\cal O}(\lambda g^2)$, ${\cal O}(\beta^2 g^2)$ terms in \ref{e66}, we find a real and positive result for $\langle \bar{\phi}^2 \rangle_{\rm loc} $, as depicted in \ref{f4} and \ref{f5}. 

\begin{figure}[h!]
\begin{center}
   \includegraphics[width=10.0cm]{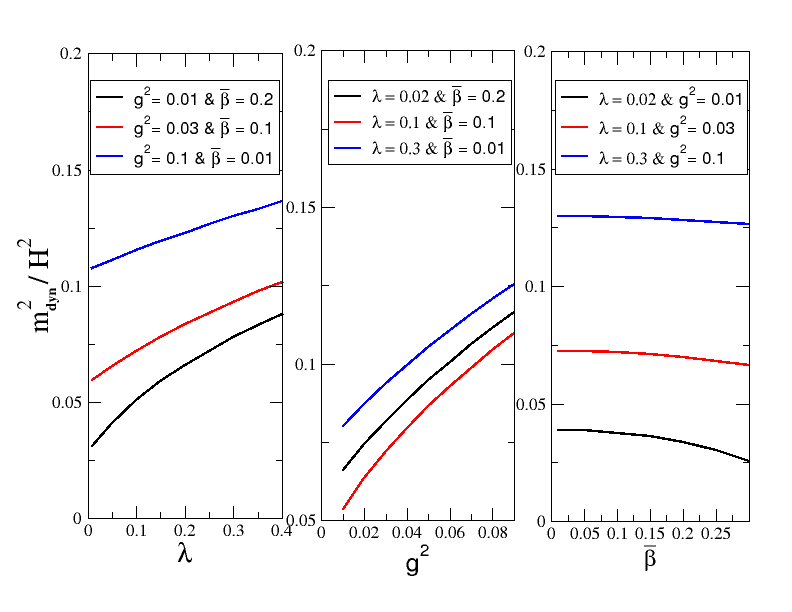}
 \caption{\small \it The variation of the dynamical mass, \ref{e52add}, of the scalar field with respect to different couplings. The decrease in $m^2_{\rm dyn}/H^2$ with the cubic coupling certainly corresponds to the destabilising effect of the cubic scalar potential.}
  \label{f6}
\end{center}
\end{figure}

A finite and non-vanishing $\langle \bar{\phi}^2 \rangle_{\rm loc} $ also indicates mass for the scalar field, dictated by the formula~\cite{davis, Beneke:2012kn}
\be
\frac{m^2_{\rm dyn}}{H^2}= \frac{3}{8\pi^2 \langle \bar{\phi}^2\rangle_{\rm loc}}
\label{e52add}
\ee
which is depicted in \ref{f6}. Since the scalar field was massless initially, this mass has generated dynamically. Since a massive scalar, no matter how tiny its mass is, obeys the de Sitter invariance, the dynamical mass generation clearly shows that the secular effect cannot eventually destroy the de Sitter spacetime. However, the backreaction due to the matter field might change the inflationary $\Lambda$ value.  We refer our reader for earlier discussion on dynamical mass generation in de Sitter spacetime for a self interacting scalar field~\cite{Serreau:2013eoa,  Youssef:2013by, Kamenshchik:2020yyn, Kamenshchik:2021tjh, Bhattacharya:2022wjl}, and also to~\cite{Prokopec:2003tm} in the context of scalar quantum electrodynamics.  \\

\noindent
Before we end, we wish to comment on $\langle \phi^2\rangle_{\rm NL}$ mentioned in the previous section. First note that if we accept that we should consider diagrams  which only generate 1PI self energy contributions for the scalar two point correlation function, we should further exclude from \ref{e49} all such non-desired terms, for example from the group \ref{f2'}. Next, since the scalar has dynamically become massive now, it seems natural that we should incorporate the same into \ref{e49},  in order to get the correct physics. For the mass term only, $\langle \phi^2 \rangle$ falls off as $e^{-m^2_{\rm dyn}{\cal N}/H^2}$. Thus as the simplest approximation, one may wish to treat the non-local terms in \ref{e49} as sources and can substitute $\langle \phi^2 \rangle \sim e^{-m^2_{\rm dyn}{\cal N}/H^2}$ there, giving $\langle \phi^2 \rangle_{\rm NL} \to 0$ at late times.  However to the best of our understanding,  in order to draw any clear conclusion on $\langle \phi^2 \rangle_{\rm NL}$, it seems that we need a more careful and rigourous analysis regarding the 1PI non-local part of the self energy. Specifically, along with the dynamically generated scalar mass, we should also consider resumming  those  self energy constributions, instead of making the aforementioned substition of $e^{-m^2_{\rm dyn}{\cal N}/H^2}$ in place of them. Perhaps we might even need to go to higher order loop computations. This seems to be a non-trivial and technically non-trivial task compared to that of the pure local contributions, and we wish to come back to this issue in a future work.

%%%%%%%%%%
\section{Discussion}\label{s5}
%%%%%%
In this work we have considered the secular effect for  a massless and minimal quantum scalar  field with an asymmetric self interaction  (\ref{f0}) coupled to massless fermion via the Yukawa interaction, \ref{y4}, in the inflationary de Sitter background.  We note that this potential has shape wise similarity with the standard slow roll inflationary ones.  The scalar is initially assumed to be located around the flat region $(\phi \sim 0)$ of the potential, so that perturbation theory with respect to the  Bunch-Davies vacuum holds good initially. As time goes on, the system should roll down towards the minima of the potential. However in the due course of this rolling, there will be strong quantum effects, dictated by the secular logarithms, originating from the highly stretched, super-Hubble late time modes.  
We have considered the late time large wavelength equation of motion for the scalar field,  and have  constructed the first order differential equations satisfied by $\langle \phi \rangle $ and $\langle \phi^2 \rangle $ (with respect to the initial Bunch-Davies vacuum), respectively \ref{e2}, \ref{e33}. These equations are sourced by vacuum expectation values of various combinations of the field operators. We have computed these vacuum expectation values up to three loop, in terms of the leading secular logarithms. Using these perturbative results,  we have constructed next the non-perturbative equations of motion for $\langle \phi \rangle $ and $\langle \phi^2 \rangle $ and have found out their non-perturbative values respectively in \ref{s3} and \ref{s4}. \ref{e30} shows that the magnitude $\langle \phi \rangle $ decreases with increasing Yukawa coupling. We also note that the resulting  $\langle \phi \rangle$ is not close to $-3\beta/\lambda$ i.e., the point of classical minimum of the potential, clearly revealing the strong quantum effect. On the other hand, a finite and non-vanishing $\langle \phi^2 \rangle $ in particular, indicates a dynamically generated mass of the scalar field, \ref{e52add}, which has been investigated in the preceding section.  On the other hand,  a detailed and systematic analysis for the resummed $\langle \phi^2 \rangle $ containing the effect of the non-local self-energy as well as the dynamical mass will be reported in a future work (cf., discussion at the end of the preceding section).  

We have seen that if we consider all the diagrams for the computation of $\langle \phi^2 \rangle$, we may obtain complex values for certain parameter ranges (cf., discussion below \ref{e66}), which is certainly unacceptable for $\phi$ is hermitian. As we have argued, since the diagrams correspond to the derivative of  $\langle \phi^2 \rangle$, and the subsequent  integration basically inserts an external point on a scalar line in any diagram, we should first check whether any diagram, after such insertion, yields an 1PI contribution. This not only rules out the non-1PI diagrams for example from the group \ref{f2'}, but also the last two diagrams of \ref{f3'} which are 1PI to begin with. Note for these two latter diagrammes that the external points after integration are to be inserted between $x$ and $x''$, and the original vertices at $x$ becomes  dummy. Accordingly, the two resulting diagrams are not 1PI. Perhaps one should keep in mind this algorithm while resumming the scalar two point function. We also note that we did not encounter any such problem while computing $\langle \phi \rangle$, for if we integrate any diagram corresponding to its derivative, we basically generate a non-amputated tadpole, \ref{fig-C}. Clearly such diagram cannot be   1PI (cf., the discussion below \ref{e66}).

It seems to be an important task to compute a non-perturbative vacuum expectation values of the interaction potential   and the scalar self interaction potential. Finding out a non-perturbative effective potential seems to be an interesting task too. We wish to come back to these issues in our future publications. 

\bigskip

%%%%%%%%%%%%%%%%
\section*{Acknowledgements} 
SB would like to acknowledge late Prof Theodore N Tomaras for useful discussions on fermions in the de Sitter spacetime. The authors would like to acknowledge anonymous referee for a careful critical reading of an earlier version of this manuscript and for making useful comments. 
%%%%%%%%%%%%%%%%%%%%%%%%%%%%%%%%%%%

\bigskip
\appendix
\labelformat{section}{Appendix #1} 
%%%%%%%%%
\section{The in-in formalism}\label{A}
%%%%%%%%%%
The standard quantum field theory involves computing $S$-matrix elements of observables with respect to in and out states while the vacuum of the theory is uniquely defined and stable. For a dynamical background such as  the de Sitter however, the initial vacuum state is not stable and decays into a final one due to particle pair creation.   Apart from this we also note that the interactions {\it cannot } be adiabatically turned on and off in a cosmological scenario and in fact they are present all the way throughout. Moreover unlike the flat spacetime, we do not have a freedom to define our initial and final states respectively at temporal past and future infinities.  In such backgrounds, one needs the Schwinger-Keldysh or the in-in or the closed time path  formalism to compute the expectation value of any operator meaningfully, which we wish to review very briefly below, referring  our reader to~\cite{Chou, Calzetta, Hu, Weinberg, Adshead} and references therein  for excellent pedagogical detail. 

The functional integral representation of the standard in-out matrix elements with respect to the field basis reads
\begin{eqnarray}\label{sw}
\langle\phi|T(O_1[\phi])| \psi\rangle=\int \mathcal{D} \phi \,e^{i \int_{t_{i}}^{t_{f}} \sqrt{-g} d^{d} x \mathcal{L}[\phi]} \Phi^{\star}\left[\phi\left(t_{f}\right)\right] O_1[\phi] \Psi\left[\phi\left(t_{i}\right)\right]
\end{eqnarray}
where $T$ stands for time ordering, $O_1[\phi]$ is some observable and $\Phi$, $\Psi$ are the wave functionals with respect to the field kets $|\phi\rangle$ and $|\psi\rangle$ respectively. We also have the representation for anti-time ordering as 
\begin{eqnarray}\label{ky}
\langle\psi|\bar{T}(O_2([\phi]))| \phi\rangle=\int \mathcal{D} \phi e^{-i \int_{t_{i}}^{t_{f}} \sqrt{-g} d^{d} x \mathcal{L}[\phi]} \Phi\left[\phi\left(t_{i}\right)\right] O_2[\phi] \Psi^{\star}\left[\phi\left(t_{f}\right)\right]
\end{eqnarray}
 Combining the above with \ref{sw}, we write down the in-in matrix representation
\begin{eqnarray}\label{ex}
\langle\psi|\bar{T}(O_2[\phi]) T(O_1[\phi])| \psi\rangle=\int \mathcal{D} \phi_{+} \mathcal{D} \phi_{-} \delta\left(\phi_{+}\left(t_{f}\right)-\phi_{-}\left(t_{f}\right)\right) e^{i \int_{t_{i}}^{t_{f}} \sqrt{-g} d^{d} x\left(\mathcal{L}\left[\phi_{+}\right]-\mathcal{L}\left[\phi_{-}\right]\right)} \Psi^{\star}\left[\phi_{-}\left(t_{i}\right)\right] O_2\left[\phi_{-}\right] O_1\left[\phi_{+}\right] \Psi\left[\phi_{+}\left(t_{i}\right)\right] \nonumber\\
\end{eqnarray}
Note that we have used two different kind of scalar fields $\phi_{\pm}$. $\phi_{+}$ evolves the system forward in time and $\phi_{-}$ evolves it backward. $\phi_{+}$  and $\phi_{+}$ are  coincident on the final hypersurface at $t=t_{f}$.  We also have used  the completeness relation on the final hypersurface 
 $$ \int \mathcal{D} \phi \,\Phi[\phi_{-}\left(t_{f}\right)] \Phi^{\star}[\phi_{+}\left(t_{f}\right)]=\delta(\phi_{+}(t_{f})-\phi_{-}(t_{f}))$$
 The field excitations corresponding to $\phi_-$ always stand for virtual particles, whereas $\phi_+$ correspond to both real and virtual excitations.
 
The Wightman functions in this formalism with respect to the initial vacuum state read
\begin{eqnarray}
i\Delta^{-+}(x,x^\prime) =
\langle \phi^-(x)\phi^+(x^\prime)\rangle \qquad \qquad
i\Delta^{+-}(x,x^\prime) =
\langle \phi^-(x')\phi^+(x)\rangle
 \label{propagators}
\end{eqnarray}
where we have taken $\eta\gtrsim \eta'$ above. On  the other hand, the Feynman and anti-Feynman propagators for the time ordered and anti-time ordered cases read respectively
\begin{eqnarray}
\label{propagatoridentities}
i\Delta^{++}(x,x^\prime) &=& \theta(\eta-\eta^\prime)i
\Delta^{-+}(x,x^\prime) + \theta(\eta^\prime-\eta)i
\Delta^{+-}(x,x^\prime) \nonumber
\\
i\Delta^{--}(x,x^\prime) &=& \theta(\eta^\prime-\eta)i
\Delta^{-+}(x,x^\prime) + \theta(\eta-\eta^\prime)i
\Delta^{+-}(x,x^\prime) 
\end{eqnarray}

The fermion can also be included in \ref{ex} in a similar manner, by introducing the fields, $\psi^{\pm}$ and the adjoints $\bar{\psi}^{\pm}$. The four fermion propagators $iS^{\pm \pm}$ can also be defined in a similar manner as that of the scalar.

%%%%%%%%%%%%%%%%%%%%%%
\section{Tree level infrared effective scalar propagators}\label{B}
%%%%%%%%%%%%
In this Appendix we shall briefly review for the sake of completeness,  the infrared (IR) truncated tree level scalar propagators necessary to compute  the late time secular effect in the  de Sitter spacetime, used in the main body of this paper.  The IR effective field theory corresponds to the super-Hubble modes at late cosmological times, immensely  stretched due to the enormous spacetime expansion.   As the name suggests, this framework is devoid of any ultraviolet divergences. Such field theory corresponds to IR truncated modes written as, e.g.~\cite{Onemli:2015pma},
\begin{eqnarray}
\phi(\eta, \vec{x}) \vert_{\rm IR} = \int \frac{d^3 \vec{k}}{(2\pi)^{3/2}} \theta (Ha-k) \left[a_{\vec k}\, u_k(\eta)\vert_{\rm IR} e^{-i\vec{k}\cdot \vec{x}}+a^{\dagger}_{\vec k}\, u^{\star}_k(\eta)\vert_{\rm IR} e^{i\vec{k}\cdot \vec{x}}\right]
\label{c1}
\end{eqnarray}
where $k=|\vec{k}|$ and  the operators satisfy the canonical commutation relation $[a_{\vec k}, a^{\dagger}_{\vec k'}]= \delta^3(\vec{k}-\vec{k'})$. The mode functions $u_k(\eta)$ for a massless and minimally coupled scalar in the de Sitter background \ref{y0}, corresponding to the Bunch-Davies vacuum is given by 
$$u_k(\eta) = \frac{H}{\sqrt{2k^3}} \left(1+ik\eta \right) e^{-ik\eta}$$
 The late time ($|H\eta| \ll 1 $), super-Hubble IR modes $(k \lesssim aH)$ can be read off as, 
\begin{eqnarray}
u_{k}(\eta) \vert_{\rm IR} \approx \frac{H}{\sqrt{2} k^{3/2}}\left[1+\frac12 \left(\frac{k}{Ha} \right)^2 + \frac{i}{3}\left(\frac{k}{Ha} \right)^3 +\,{\rm subleading~terms } \right]
\label{c2}
\end{eqnarray}
Thus the temporal part of the Bunch-Davies modes approaches  nearly a constant in this limit. Substituting this into \ref{c1} and dropping  the suffix `IR' without any loss of generality, we have
\begin{eqnarray}
\phi(\eta, \vec{x})  =\frac{H}{\sqrt{2}} \int \frac{d^3 \vec{k}}{(2\pi)^{3/2}} \frac{\theta (Ha-k)}{k^{3/2}} \left[a_{\vec k}\,  e^{-i\vec{k}\cdot \vec{x}}+a^{\dagger}_{\vec k}\,  e^{i\vec{k}\cdot \vec{x}}\right]+\,{\rm subleading~terms}
\label{c3}
\end{eqnarray}
Using the above and $H \lesssim  k \lesssim Ha$ for these mode functions, it is straightforward to compute 
\be
\langle \phi^2(x) \rangle=\frac{H^2}{4\pi^2} \ln a 
\label{free}
\ee
which corroborates with the renormalised version of \ref{y6}. As expected in an IR effective field theory, the coincidence limit expression \ref{free} is free from any ultraviolet divergence.

We also note the canonical momentum conjugate to the IR scalar \ref{c3},
\begin{eqnarray}
 \dot{\phi}(\eta, \vec{x})= H^2 \int \frac{d^3 \vec{k}}{(2\pi)^{3/2}} \frac{\delta(k-Ha)}{\sqrt{2k}}\left[a_{\vec k}\,  e^{-i\vec{k}\cdot \vec{x}}+a^{\dagger}_{\vec k}\,  e^{i\vec{k}\cdot \vec{x}}\right]
\label{c7}
\end{eqnarray}
where the dot denotes differentiation once with respect to the cosmological 
time. It is now easy to see from  \ref{c3} that $[\phi(t,\vec{x}), \dot{\phi}(t,\vec{x}')]_{\rm IR}\approx 0$. This is a manifestation of the fact that the IR field by and large is no longer quantum, but is stochastic.   \\

\noindent
Finally, we note the propagators necessary for the in-in computations, \ref{ex}. Defining the spatial Fourier transform,
\begin{eqnarray}
i\Delta(x,x')
= \int \frac{d^3 {\vec k}}{(2\pi)^3} e^{i{\vec k}\cdot({\vec x}-{\vec y})}i \Delta(k,\eta,\eta'),
\label{nc2}
\end{eqnarray}
we have from \ref{propagators} at the leading order
\begin{eqnarray}
 i \Delta_{+-} (k,\eta,\eta')&&\approx  \frac{H^2 \theta(Ha-k)\theta(Ha'-k)}{2k^3} \left(1+\frac{ik^3}{3H^3 a'^3} \right)\qquad (\eta \gtrsim \eta')\nonumber\\
&& \approx \frac{H^2 \theta(Ha'-k)\theta(Ha-k)}{2k^3} \left(1-\frac{ik^3}{3H^3 a^3} \right)\qquad (\eta' \gtrsim \eta)\label{nc4'}
\label{nc4'}
\end{eqnarray}
We also note for our purpose
\begin{eqnarray}
&& i \Delta_{+-} (k,\eta,\eta') - i \Delta_{-+} (k,\eta,\eta')\approx 
\frac{i \theta(Ha-k)\theta(Ha'-k)}{3 H a'^3}  \nonumber\\
&&i \Delta_{+-} (k,\eta,\eta') + i \Delta_{-+} (k,\eta,\eta')\approx \frac{H^2}{k^3}\theta(Ha-k)\theta(Ha'-k) \qquad (\eta \gtrsim \eta')
\label{nc4}
\end{eqnarray}

The upper limit of the momenta inside any loop will naturally be decided by the step functions appearing above. The lower limit will be taken to be $H$ in our  IR effective description, in order to account for the super-Hubble modes~\cite{Onemli:2015pma}.  Also in the main text,  the temporal argument  $\eta$ will be regarded as the final time and any other dummy temporal variable inside the loop will be at most taken to be equal to $\eta$. 

%%%%%%%%%%%%%%%%%%%%%%

%%%%%%%%%%%%%%%%%%%%%%
\section{A consistency check for computation of $\langle \phi \rangle$, \ref{s3}}\label{C}
%%%%%%%%%%%%

%
\begin{figure}[h!]
\begin{center}
  \includegraphics[width=15.0cm]{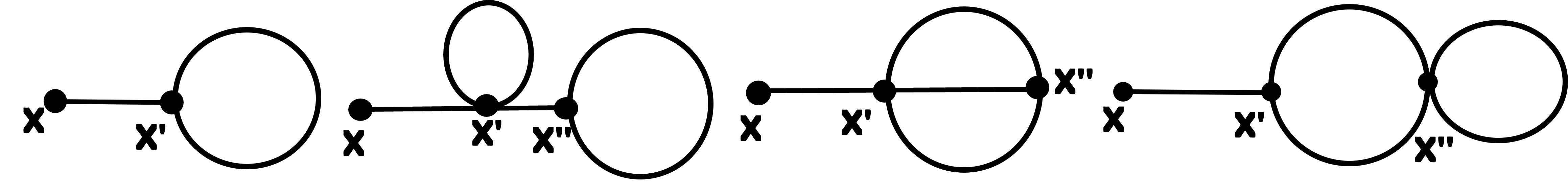}
 \caption{\small \it One and two loop tadpoles for the scalar self interaction to computate $\langle \phi (x)\rangle$ perturbatively.  All primed vertices  are integrated over. }
  \label{fig-C}
\end{center}
\end{figure}
\noindent
In this appendix we wish to show for the sake of consistency that the perturbative expression for $\langle \phi \rangle $ derived in \ref{s3} essentially equals the one computed directly from the one and two loop tadpoles,  \ref{fig-C}. We shall use the in-in formalism and infrared effective propagators outlined in \ref{A}, \ref{B}. It will be sufficient to focus only on the scalar sector. Computation of the same using the exact propagator and subsequent renormalisation can be seen in~\cite{Bhattacharya:2022aqi}.

At one loop (the first of \ref{fig-C}), we have in our IR effective formalism   
\begin{eqnarray}
\langle \phi \rangle = \frac{i\beta}{2} \int d^4 x' a'^4 i\Delta (x',x') \left(i\Delta_{+-} (x,x')-i\Delta_{-+}(x,x') \right)
\label{C1}
\end{eqnarray}
Introducing now the spatial momentum space as of \ref{B}, using \ref{free} ($i\Delta (x,x)=\langle \phi^2\rangle$) and \ref{nc4}, the above becomes 
\begin{eqnarray}
\langle \phi \rangle = \frac{i\beta H^2}{2^3 \pi^2} \int d \eta' a'^4 \ln a' \left(i\Delta_{+-} (0,\eta, \eta')-i\Delta_{-+}(0,\eta, \eta') \right) = -\frac{\beta}{2^3\times 3\pi^2}\int \frac{da'}{a'}\ln a'= -\frac{\beta {\cal N}^2}{2^4\times 3\pi^2}
\label{C2}
\end{eqnarray}
where ${\cal N}=Ht$ as earlier.  This exactly equals with the one obtained from the first diagram of \ref{f21}, and subsequently multiplied \ref{e14-} by $-\beta/6H$ and then integrating it with respect to  ${\cal N}$, giving the ${\cal O}(\bar \beta)$ term on the right hand side of \ref{e26}.   \\

\noindent
The second diagram of \ref{fig-C} is two loop, reading
\begin{eqnarray}
\langle \phi \rangle = &&-\frac{\lambda \beta}{2^2} \int d^4 x' d^4 x'' (a'a'')^4 i\Delta(x',x') i\Delta(x'',x'') \nonumber\\
&&\times\left[i\Delta_{++}(x,x')\left(i\Delta_{++}(x',x'')- i\Delta_{+-}(x'x'')\right)+ i\Delta_{+-}(x,x')\left( i\Delta_{--}(x',x'')- i\Delta_{-+}(x',x'')\right) \right] \nonumber\\
\label{C3}
\end{eqnarray}
Note that $\eta$ is the final time. Also, from \ref{propagatoridentities}, we see that the above integral is non-vanishing only for the temporal hierarchy, $\eta \gtrsim \eta' \gtrsim \eta''$.  Using this and the spatial momentum space and using \ref{nc4}, we have 
\begin{eqnarray}
&&\langle \phi \rangle = -\frac{\lambda \beta H^4}{2^6\pi^4} \int d^4 x' d^4 x'' (a'a'')^4 \ln a' \ln a'' 
\left(i\Delta_{+-}(x,x')-i\Delta_{-+}(x,x')\right)\left(i\Delta_{+-}(x',x'')- i\Delta_{-+}(x',x'')\right)\nonumber\\
&&=-\frac{\lambda \beta H^4}{2^6\pi^4} \int d \eta' d \eta'' (a'a'')^4 \ln a' \ln a'' 
\left(i\Delta_{+-}(0,\eta,\eta')-i\Delta_{-+}(0,\eta,\eta')\right)\left(i\Delta_{+-}(0,\eta',\eta'')- i\Delta_{-+}(0,\eta',\eta'')\right)
\nonumber\\
&&=\frac{\lambda \beta }{2^6\times 9\pi^4} \int \frac{da'}{a'}\ln a' \int \frac{da''}{a''}\ln a''= \frac{\lambda \beta {\cal N}^4}{2^9\times 9\pi^4} 
\label{C4}
\end{eqnarray}
The above exactly matches with the one obtained from the first of \ref{f2'} (\ref{e4}), multiplied with $-\lambda/18H$, and then integrating it with respect to ${\cal N}$.

The third diagram of \ref{fig-C} equals 
\begin{eqnarray}
&&-\frac{\lambda \beta}{6} \int d^4 x' d^4 x'' (a'a'')^4  \left[ i\Delta_{++}(x,x') \left(i\Delta^3_{++}(x',x'')- i\Delta^3_{+-}(x',x'')\right)+  i\Delta_{+-}(x,x') \left(i\Delta^3_{--}(x',x'')- i\Delta^3_{-+}(x',x'')\right)\right]\nonumber\\
\label{C5'}
\end{eqnarray}
Thus the above diagram is non-vanishing for the temporal hierarchy, $\eta \gtrsim \eta'\gtrsim \eta''$. Employing the spatial momentum space, and using then \ref{nc4'}, \ref{nc4}, the above integral is evaluated as
\begin{eqnarray}
&&-\frac{\lambda \beta}{6} \int d^4 x' d^4 x'' (a'a'')^4 \left(i\Delta_{+-}(x,x')- i\Delta_{-+}(x,x')\right) 
\left(i\Delta^3_{+-}(x',x'') - i\Delta^3_{-+}(x',x'')\right)\nonumber\\
&&= -\frac{\lambda \beta}{6} \int d\eta' d\eta'' \frac{d^3 \vec{k_1} d^3 \vec{k_2}}{(2\pi)^6}  \left( i\Delta_{+-}(0,\eta,\eta')- i\Delta_{-+}(0,\eta,\eta')\right)\left(i\Delta_{+-}(k_1,\eta',\eta'')i\Delta_{+-}(k_2,\eta',\eta'')i\Delta_{+-}(|\vec{k_1}+\vec{k_2}|,\eta',\eta'') - {\rm c.c.}\right)\nonumber\\
&&= \frac{\lambda \beta {\cal N}^4}{2^7 \times 27\pi^4}
\label{C5}
\end{eqnarray}
where `c.c.' denotes complex conjugation. The last expression of the above equation exactly equals the integral of the second diagram of \ref{f2'} (\ref{e6}) multiplied with $-\lambda/18H$, as earlier. \\

\noindent
Finally, we wish to evaluate the last diagram of \ref{fig-C} in a manner similar to the above, 
\begin{eqnarray}
\langle \phi \rangle && =-\frac{\lambda \beta}{4} \int d^4 x' d^4 x'' (a'a'')^4 i\Delta(x'',x'')\left[ i\Delta_{++}(x,x') \left(i\Delta^2_{++}(x',x'')- i\Delta^2_{+-}(x',x'')\right)\right. \nonumber\\&&\left. +  i\Delta_{+-}(x,x') \left(i\Delta^2_{--}(x',x'')- i\Delta^2_{-+}(x',x'')\right)\right] = \frac{\lambda \beta {\cal N}^4}{2^7 \times 27\pi^4}
\label{C6}
\end{eqnarray}
which equals the one obtained by integrating the second diagram of \ref{f21} (\ref{e14}), as earlier.

If we combine now \ref{C2}, \ref{C4}, \ref{C5} and \ref{C6}, the resulting expression exactly matches with the scalar part (i.e., $g=0$) of \ref{e26}. This shows a consistency between the approach of \ref{c3} and the direct computation of $\langle \phi\rangle$.

%%%%%%%%%
\bigskip
\bigskip
%%%%%%%%%%%%%%%%%%%%%%%%%%%%%%%%%%%%%%%%%%%%%%%%%%%%%%%%%%%%%%%%%%%%%%%%%%%

\end{document}